\documentclass[aps,prb,reprint,groupedaddress,showpacs]{revtex4-1}

\usepackage{graphicx}
\usepackage{tikz}
\usepackage[caption=false]{subfig}
\usepackage{amsmath}
\usepackage{amssymb}
\usepackage{braket}
\usepackage{xfrac}
\usepackage{url}
\usepackage{filecontents}

\makeatletter
\newcommand{\colorcaption}[2][]{%
  \begingroup%
  \renewcommand{\@caption@fignum@sep}{ (Color online). }%
  \caption[#1]{#2}%
  \endgroup%
}
\makeatother

\renewcommand{\vec}[1]{\boldsymbol{#1}}
\newcommand{\efield}{\vec{\mathcal{E}}}
\newcommand\myatop[2]{\genfrac{}{}{0pt}{}{#1}{#2}}
\newcommand{\BTO}{$\mathrm{BaTiO_3}$}
\newcommand{\PTO}{$\mathrm{PbTiO_3}$}

\begin{document}

\title{First-principles calculations for ferroelectrics at constant polarization\\using generalized Wannier functions}

\author{Pawel Lenarczyk}
\email{pawell@iis.ee.ethz.ch}
\author{Mathieu Luisier}
\affiliation{Integrated Systems Laboratory, ETH Z\"{u}rich, 8092 Z\"{u}rich, Switzerland}

\date{\today}

\begin{abstract}
Localized Wannier functions provide an efficient and intuitive framework to compute electric polarization
from first-principles. They can also be used to represent the electronic systems at fixed electric field
and to determine dielectric properties of insulating materials.
Here we develop a Wannier-function-based formalism to perform first-principles calculations at fixed
polarization. Such~an~approach allows to extract the polarization-energy landscape of a crystal and
thus supports the theoretical investigation of polar materials.
To facilitate the calculations, we implement a~\mbox{quasi-Newton} method that simultaneously relaxes the internal
coordinates and adjusts the electric field in crystals at fixed polarization.
The method is applied to study the ferroelectric behavior of $\mathrm{BaTiO_3}$ and $\mathrm{PbTiO_3}$
in tetragonal phases. The physical processes driving the ferroelectricity of~both compounds are examined
thanks to the localized orbital picture offered by Wannier functions. Hence, changes in chemical bonding
under ferroelectric distortion can be accurately visualized. The~difference in the ferroelectric properties of
$\mathrm{BaTiO_3}$ and  $\mathrm{PbTiO_3}$ is highlighted. It can be traced back to the peculiarities of
their electronic structures.
\end{abstract}


\maketitle

\section{Introduction\label{sec_intro}}

The development of the microscopic modern theory of polarization \cite{ref_MTP_KSV_I,ref_MTP_Resta,ref_MTP_KSV_II}
(MTP) has enabled significant progresses in the understanding of ferroelectric states. MTP rigorously defines the
polarization of a periodic solid and provides a route for its computation using electronic structure methods
such as density functional theory \cite{ref_DFT} (DFT). Thus, many properties that could previously be
inferred only at a very qualitative level can now be computed with quantum mechanical accuracy
from first-principles.

MTP is also a basis for the development of techniques to determine the exact ground state of a crystalline insulator
in the presence of an electric field $\efield$. This is realized \cite{ref_NV,ref_SIV,ref_MD_efield} by minimizing
the electric enthalpy functional composed of the usual Kohn-Sham energy and a field coupling term:
\mbox{``$\efield \cdot \vec{P}$''}, involving the electric polarization $\vec{P}$.
In particular, it has been shown by the authors in Ref.~\onlinecite{ref_my_NGWF} that non-orthogonal generalized Wannier functions
provide efficient and intuitive means by which to perform finite-field calculations within a DFT framework.
It is therefore possible to calculate from first-principles many interesting properties of materials
related to their behavior under external electric fields.

In this paper we present a method for performing first-principles calculations not at constant
electric field, but at fixed electric polarization.
This enables to compute crystal properties as a function of $\vec{P}$, providing a way
to extract the polarization-energy landscape~$E(\vec{P})$ of any material.
Knowing $E(\vec{P})$, its dielectric or ferroelectric properties can be inferred.
Moreover, constrained-$\vec{P}$ calculations make it simultaneously possible to exhibit and understand
the dynamical transformations of the system under study, which can lead to a particular electrical behavior.
In addition to first-principles investigations, the ability to compute $E(\vec{P})$ within DFT provides an
intuitive link to Landau-Devonshire (LD) semiempirical theory \cite{ref_LandauPrimer} in which $\vec{P}$
serves as an order parameter. Hence, the \mbox{constrained-$\vec{P}$} method may pave the way for the
first-principles derivation of LD descriptions for a wide range of ferroelectric materials.
This could be useful, for example, in the context of ferroelectric device simulations.\cite{ref_myFeFET}

Our approach partly derives inspiration from the work of Sai, Rabe, and Vanderbilt (SRV) on the structural response
to macroscopic electric fields in ferroelectric systems.\cite{ref_SRV} By~using density functional perturbation
theory (DFPT) these authors constructed an approximate thermodynamic potential whose minimization with respect to
internal structural parameters produces crystal structures at fixed polarization. In the SRV approach both the
internal energy and the electric polarization are evaluated at zero electric field. The effect of the
electric field on the electronic structure is therefore neglected. This limitation of the SRV method was addressed by
Dieguez and Vanderbilt \cite{ref_DV} (DV) who overcome it by employing the theory of finite electric fields developed by
Souza, Iniguez, and Vanderbilt \cite{ref_SIV} (SIV). The DV method is thus an exact one. However, because it incorporates
the reciprocal-space-based SIV theory of finite electric fields, Brillouin zone sampling may be restricted
to eliminate the possibility of runaway solutions,\cite{ref_SIV,ref_NG} i.e., to allow for stable
stationary solutions to exist. This is especially problematic when it comes to the
calculation of ferroelectric properties, which are known to be sensitive to the quality of the Brillouin zone integration
and require large \mbox{$k$-point} sets to converge.\cite{ref_Cohen_BTO_PTO,ref_BTO_PP}

Instead, we base our constrained-$\vec{P}$ method on the first-principles theory of finite electric fields proposed by
Nunes and Vanderbilt \cite{ref_NV} (NV). These authors showed in Ref.~\onlinecite{ref_NV} that a real-space representation
of occupied subspace in terms of orthogonal Wannier-functions (WFs) could be used to calculate the internal energy and
the electric polarization of a periodic insulator in the presence of a finite electric field.
In particular, we rely on our extension of this theory to employ non-orthogonal generalized Wannier functions (NGWFs),
as implemented in Ref.~\onlinecite{ref_my_NGWF}.
The purpose is twofold: apart from its computational efficiency, such a formulation readily allows for an intuitive
understanding of the effects of the polarization. As will become clear below, this is possible by employing a Wannier-like
representation of the electronic structure. Moreover, this description provides an insightful picture of the nature of the
chemical bonds in materials,\cite{ref_MLWF} otherwise missing from the picture of extended eigenstates: insightful chemical
analyses of the nature of bonding can be performed, as well as its evolution during ferroelectric transitions. Finally, the
mechanism of polarization as electronic currents produced by dynamical changes of orbital hybridizations can be visualized,
helping to clarify the origin of ferroelectricity in polar materials.

The plan of the remainder of this paper is as follows.
In the next section, the details of the general theoretical framework are presented.
The practical implementation of the method is then discussed in Sec.~\ref{sec_calc}.
In~Sec.~\ref{sec_alg} a description of the quasi-Newton algorithm for constrained polarization
calculations is given. The systems analyzed in this work are described in Sec.~\ref{sec_comput},
including a discussion of the technical details of the \emph{ab initio} pseudopotential calculations.
Then, in Sec.~\ref{sec_results}, we report the results of our calculations, starting in Sec.~\ref{sec_conv}
with the tests that have been performed to probe the practical usefulness of the method.
The method is then applied in Sec.~\ref{sec_BTO} and \ref{sec_PTO} to study the ferroelectricity of tetragonal
\BTO{} and \PTO{}, respectively. The $E(\vec{P})$ characteristics are obtained and associated with dynamical
transformations of the crystal and electronic structures at fixed polarization. Localized Wannier functions
are used to investigate the changes in chemical bonding that lead to the stabilization of a ferroelectric distortion
for both compounds. We also highlight the differences in the ferroelectric behavior of \BTO{} and \PTO{} that
come from their different electronic structures.
Finally, in Sec.~\ref{sec_concl}, we summarize and conclude our work.

Atomic (Rydberg) units are used throughout this paper, unless explicitly stated, i.e., $\bar{e}=\hbar=m_{e}=1$,
length unit $\mathrm{r_B}=0.53\mathrm{\AA}$, energy unit $\mathrm{Ry}=13.6\mathrm{eV}$.

\section{Formalism\label{sec_formalism}}
We will now describe a formalism to perform first-principles calculations at constant polarization.
Our approach replicates some of the ideas of the SIV perturbative scheme,\cite{ref_SRV} but results instead in an exact method
since it incorporates the NV theory of finite electric~fields.\cite{ref_NV} Moreover, in the derivation presented below
forces due to polarization constraint appear in a more natural way than in the SIV approach.

Let $\rho$ be the electron density and $\vec{\tau}$ the atomic coordinates.
Consider a periodic insulating crystal with a unit cell volume $\Omega$.
In the context of electronic structure calculations, the electric enthalpy is given by \cite{ref_NV,ref_NG,ref_SIV,ref_my_NGWF,ref_MD_efield}
\begin{equation}
 W[\rho](\vec{\tau},\efield) = E_{\mathrm{KS}}[\rho](\vec{\tau}) - \Omega \efield \cdot \vec{P}[\rho](\vec{\tau}) ~.
 \label{eq_enthalpy}
\end{equation}
Here $E_{\mathrm{KS}}[\rho](\vec{\tau})$ stands for the usual Kohn-Sham (KS) energy per unit cell.
The above equation must be minimized with respect to the density functions $\rho$ in order to find the electronic
ground state in the presence of an electric field
\begin{equation}
 \begin{split}
  W_{\mathrm{gs}}(\vec{\tau},\efield) & = \min_{ \small { \rho } } W[\rho](\vec{\tau},\efield) \\
                                      & = E_{\mathrm{KS}}(\vec{\tau}) - \Omega \efield \cdot \vec{P}(\vec{\tau}) ~.
 \end{split}
 \label{eq_Egs}
\end{equation}
This solution corresponds to a polarized long-lived resonant state of a periodic insulating solid, where the intraband (or Zener) tunneling
is neglected.\cite{ref_NV,ref_NG,ref_SIV} The functional $W_{\mathrm{gs}}(\vec{\tau},\efield)$ can be viewed as a thermodynamic potential
that minimizes to equilibrium values the coordinates $\vec{\tau}$ at~fixed~$\efield$.

An approach to solve this finite-field problem, given by the minimization of the electric enthalpy Eq.~(\ref{eq_Egs}) with respect
to the field-dependent density functions, was proposed by NV in Ref.~\onlinecite{ref_NV}. In the NV theory of finite-electric fields
a representation of the density functions in terms of truncated field-polarized Wannier functions is employed to write a functional
for the band-structure energy and the electronic polarization of a solid in a uniform electric field. This scheme has been implemented
in a DFT framework in Ref.~\onlinecite{ref_my_NGWF}, where an overview of the method is given in Sec.~2, for \emph{ab initio} force
calculations in the presence of electric~fields.

Our goal in this work is to perform calculations at a given polarization target value $\vec{P}_t$. This can be achieved via a
Legendre transformation of the electric enthalpy $W_{\mathrm{gs}}(\vec{\tau},\efield)$ to a thermodynamic potential
$E_{\mathrm{gs}}(\vec{\tau},\vec{P}_t)$ in which the polarization is the control parameter and the electric field the state variable
\begin{equation}
 E_{\mathrm{gs}}(\vec{\tau},\vec{P}_t) = \min_{ \small { \efield } } \left\{ W_{\mathrm{gs}}(\vec{\tau},\efield) + \Omega \efield \cdot \vec{P}_t \right\} ~.
 \label{eq_E_tau_pol}
\end{equation}

The thermodynamic potential $E_{\mathrm{gs}}(\vec{\tau},\vec{P}_t)$ can be minimized with respect to the atomic coordinates
$\vec{\tau}$, which leads to the energy $E(\vec{P}_t)$, a functional of polarization
\begin{equation}
 \begin{split}
  E(\vec{P}_t) & = \min_{ \small { \vec{\tau} } } E_{\mathrm{gs}}(\vec{\tau},\vec{P}_t) \\
               & = \min_{ \small { \vec{\tau}, \efield } } \left\{ E_{\mathrm{KS}}(\vec{\tau}) - \Omega \efield \cdot \left( \vec{P}(\vec{\tau}) - \vec{P}_t \right) \right\} ~.
 \end{split}
 \label{eq_Epol}
\end{equation}
This minimization allows to determine the structural properties of the system under study at a fixed polarization $\vec{P}_t$.
As it can be seen, the expression in Eq.~(\ref{eq_Epol}) can also be interpreted as one where $\efield$ represents a
Lagrange multiplier implementing the constraint $\vec{P}(\vec{\tau})=\vec{P}_t$.

The governing equations of the constrained polarization calculations can be derived by applying the variational principle.\cite{ref_Zienkiewicz_variational}
Consider a change of the objective function in Eq.~(\ref{eq_Epol}), induced by a variation of the $\vec{\tau}$ and $\efield$
variables, what is equal to zero at the minimum
\begin{equation}
 \left. \frac{\mathrm{d} E_{\mathrm{gs}}(\vec{\tau},\efield)}{\mathrm{d} \vec{\tau}} \right|_{\efield} \cdot \delta \vec{\tau}
 - \Omega \left( \vec{P}(\vec{\tau}) - \vec{P}_t \right) \cdot \delta \efield = 0 ~.
 \label{eq_dEpol}
\end{equation}

By using the fact that the system is brought to the Born-Oppenheimer surface after the electronic minimization in Eq.~(\ref{eq_Egs}),
the Hellmann-Feynman theorem \cite{ref_HF_Hellmann,ref_HF_Feynman} can be invoked. This allows to take into account only the
explicit dependence of the electric enthalpy on the atomic coordinates when calculating the corresponding derivatives
\begin{equation}
 \left. \frac{\mathrm{d} W_{\mathrm{gs}}(\vec{\tau},\efield)}{\mathrm{d} \vec{\tau}} \right|_{\efield} =
 \frac{\partial E_{\mathrm{KS}}(\vec{\tau})}{\partial \vec{\tau}}
 - \Omega \efield \frac{\partial \vec{P}(\vec{\tau})}{\partial \vec{\tau}} ~.
 \label{eq_dEgs_HF}
\end{equation}
In the above equation, the first term on the right-hand side is just the negative of the force, as calculated in ordinary KS theory
\begin{equation}
 \vec{F}_{\mathrm{KS}}(\vec{\tau}) = - \frac{\partial E_{\mathrm{KS}}(\vec{\tau})}{\partial \vec{\tau}} ~.
 \label{eq_frc_KS}
\end{equation}
The second term is the force due to the polarization constraint
\begin{equation}
 \vec{F}_{\mathcal{E}}(\vec{\tau},\efield) = \Omega \efield \frac{\partial \vec{P}(\vec{\tau})}{\partial \vec{\tau}} ~.
 \label{eq_frc_efield}
\end{equation}
Therefore, it can be seen that by using the variational principle the total force
\begin{equation}
 \vec{F}(\vec{\tau},\efield) = \vec{F}_{\mathrm{KS}}(\vec{\tau}) + \vec{F}_{\mathcal{E}}(\vec{\tau},\efield) ~,
 \label{eq_frc_tot}
\end{equation}
which is composed of a KS and electric field contribution, naturally emerges in our formulation.
Practical implementation of both force terms within Wannier-function theory of finite-electric fields
will be discussed in Sec.~\ref{sec_calc}.

Since Eq.~(\ref{eq_dEpol}) must hold for arbitrary $\delta \vec{\tau}$ and $\delta \efield$ it gives after substituting Eqs.~(\ref{eq_dEgs_HF})--(\ref{eq_frc_tot})
the following system of equations
\begin{equation}
 \begin{cases}
  \vec{F}(\vec{\tau},\efield) = \vec{0} \\
  \vec{P}(\vec{\tau}) - \vec{P}_t = \vec{0}
 \end{cases}
 ~,
 \label{eq_frc_pol}
\end{equation}
which can be solved for $\vec{\tau}$ and $\efield$.
Note that Eq.~(\ref{eq_frc_pol}) corresponds to the Euler-Lagrange equation \cite{ref_Zienkiewicz_variational} of the $E(\vec{P}_t)$ functional.
The developed algorithm to solve Eq.~(\ref{eq_frc_pol}) and subsequently find the stationary point of $E(\vec{P}_t)$
will be presented in Sec.~\ref{sec_alg}.

As it can be seen from the coupled system of equations~(\ref{eq_frc_pol}), the first equation requires that the total forces
acting on the atoms vanish and the second equation ensures that the polarization constraint is fulfilled, at the end
of the geometry optimization. As it can be deduced from Eq.~(\ref{eq_Epol}), after solving Eq.~(\ref{eq_frc_pol}),
we find
\begin{equation}
 E(\vec{P}_t) = E_{\mathrm{KS}}(\vec{\tau}(\vec{P}_t)) ~.
\end{equation}
Consequently, by solving Eq.~(\ref{eq_frc_pol}) the potential energy landscape of a solid can be examined
as a function of its polarization.

\section{Calculation of Total Energy, Polarization, and Forces\label{sec_calc}}
It is not entirely straightforward to compute the ground state of a bulk solid in the presence of
an electric field, as required by Eq.~(\ref{eq_Egs}). The difficulty of treating finite electric fields is related to the
definition of a polarization for a periodic system.\cite{ref_MTP_guide} The development of the modern theory of polarization \cite{ref_MTP_KSV_I,ref_MTP_Resta,ref_MTP_KSV_II}
has been at the origin of significant recent progresses with that respect and has enabled the application of electric fields
in periodic electronic structure calculations.\cite{ref_NV,ref_SIV,ref_MD_efield} In particular, NV have shown that localized
WFs can be used to describe a periodic insulating solid in the presence of a uniform electric field.\cite{ref_NV}

The NV theory of finite electric fields and electronic polarization can be extended to the case of
non-orthogonal generalized WFs (NGWFs) to improve convergence in practical calculations.\cite{ref_my_NGWF}
The latter approach relies on the following parametrization of the density operator
\begin{equation}
 \hat{\rho} = \sum_{\myatop{ij}{ab}} \ket{\nu_a^i} K_{ab}^{ij} \bra{\nu_b^j} ~,
 \label{eq_densop_param}
\end{equation}
written in terms of the localized, Wannier-like orbitals $\{\nu_a^i\}$ and the elements of the density kernel matrix $\{K_{ab}^{ij}\}$,
where the superscript indices $i$,$j$ denote the cell replicas, and subscript indices $a$,$b$ the occupied bands.
The periodicity of the electronic state is expressed as $\ket{\nu_a^i} = \hat{T}_{\vec{R}_i} \ket{\nu_a^0}$, where
$\hat{T}_{\vec{R}_i}$ is the translation operator corresponding to the lattice vector $\vec{R}_i$ and the superscript $0$ indicates that the orbital
is centered in the unit cell containing the origin. For the density kernel matrix it holds $K_{ab}^{ij}=K_{ab}(\vec{R}_j - \vec{R}_i)$.\cite{ref_MLWF}

By taking Eq.~(\ref{eq_densop_param}) as an ansatz for a trial density
operator, the physical density operator $\hat{\rho}'$ is derived by employing the McWeeny purifying transformation.\cite{ref_McWeeny} This ensures
a weak idempotency of $\hat{\rho}'$ so that it can be used to describe the occupied space. In this approach the expectation
value of any operator $\hat{O}$  is given by $\mathrm{tr}[\hat{\rho}' \hat{O}]$.
The trace over the occupied space can be conveniently evaluated by introducing the auxiliary wave functions
\begin{equation}
 \ket{\tilde\nu_b^j} = \sum_{\myatop{i}{a}} Q_{ba}^{ji} \ket{\nu_a^i} ~,
 \label{eq_wc}
\end{equation}
where $Q_{ab}^{ij} = 2 K_{ab}^{ij} - \left( \mathbf{K} \times \mathbf{S} \times \mathbf{K} \right)_{ab}^{ij}$
and $S_{ab}^{ij} = \braket{\nu_a^i | \nu_b^j}$, the overlap matrix between the orbitals.
For the derivation of the $\mathbf{Q}$ matrix, see Ref.~\onlinecite{ref_my_NGWF}.
The $\{\tilde\nu_b^j\}$ set constitutes the biorthogonal complement \cite{ref_ON_nonorthogonal} to $\{\nu_a^i\}$,
which satisfies the biorthogonality relationship $\braket{\nu_a^i | \tilde\nu_b^j} = \delta_{ij}\delta_{ab}$.

In this formulation the band structure energy $E_{bs}$, which corresponds to the expectation value of the
Hamiltonian operator $\hat{H}$, is evaluated as
\begin{equation}
 E_{bs}[\hat{\rho}'](\vec{\tau}) = 2 \sum_{a} \bra{\nu_a} \hat{H}(\vec{\tau}) \ket{\tilde\nu_a} ~.
 \label{eq_Ebs}
\end{equation}

The electronic contribution $\vec{P}_{el}$ to the macroscopic polarization $\vec{P}$ is related to the expectation value of the
position operator,\cite{ref_PosOp} $\hat{\vec{r}}$. In the present formalism it can be simply calculated as the sum of the 
centroids of charge of the localized orbitals, scaled by the volume of the unit cell~$\Omega$
\begin{equation}
 \vec{P}_{el}[\hat{\rho}'] = - \frac{2}{\Omega} \sum_{a} \bra{\nu_a} \hat{\vec{r}} \ket{\tilde\nu_a} ~.
 \label{eq_Pel}
\end{equation}
Note that the above equation is equivalent to the expression for the electronic contribution to the polarization in the
Berry-phase theory, through the formal connections between the centers of charge of the WFs and the Berry phases of the
Bloch functions as they are carried around the Brillouin zone.\cite{ref_PosOp}

The electron density is the diagonal of the physical density matrix. In the present formalism it can be evaluated as
\begin{equation}
 \rho(\vec{r}) = 2 \sum_{a} \braket{\nu_a|\vec{r}} \braket{\vec{r}|\tilde\nu_a} ~,
 \label{eq_densval}
\end{equation}
where the factor $2$ takes into account the assumed spin degeneracy.
In Eqs.~(\ref{eq_Ebs})--(\ref{eq_densval}) and in the following, the superscript indicating the cell replica
has been dropped so that ${\nu_a \equiv \nu_a^0}$.

The Kohn-Sham total energy $E_{KS}$ can now be written in terms of the degrees of freedom of the density,
as a sum of the band-structure energy $E_{bs}$ in Eq.~(\ref{eq_Ebs}), minus the double-count correction term $E_{dc}$
and plus the classical electrostatic energy among the ions $E_{i-i}$, i.e.
\begin{equation}
 E_{KS}[\hat{\rho}'](\vec{\tau}) = E_{bs}[\hat{\rho}'](\vec{\tau}) - E_{dc}[\hat{\rho}'] + E_{i-i}(\vec{\tau}) ~.
 \label{eq_Eks}
\end{equation}

To obtain the total polarization, the classical ionic contribution $\vec{P}_{ion}$ must be added to Eq.~(\ref{eq_Pel}).
The total polarization $\vec{P}$ is then
\begin{equation}
 \vec{P}[\hat{\rho}'](\vec{\tau}) = \vec{P}_{el}[\hat{\rho}'] + \vec{P}_{ion}(\vec{\tau}) ~.
 \label{eq_Ptot}
\end{equation}
In the pseudopotential approximation employed in this work the $\vec{P}_{ion}$ term arises from the sum of the ionic core point charges
$Q_I$ located at the corresponding atomic positions $\vec{\tau}_I$
\begin{equation}
 \vec{P}_{ion}(\vec{\tau}) = \frac{1}{\Omega} \sum_{I} Q_I \vec{\tau}_I ~.
 \label{eq_Pion}
\end{equation}

By substituting Eqs.~(\ref{eq_Eks}) and (\ref{eq_Ptot}) into Eq.~(\ref{eq_enthalpy}) the minimization with respect to the degrees of freedom of the
physical density matrix, $\{\nu_a^0\}$ and $\{K_{ab}^{0i}\}$, can be carried out to determine the ground state of a solid in the presence of a
fixed electric field. In our implementation of the above outlined formalism the localized orbitals are represented on a uniform real-space grid and
$\braket{\vec{r} | \nu_a^0} $ is allowed to be nonzero only inside cubic regions with size $a_{LR}$, referred to as the localization regions (LRs)
in the following.

At the end of the electronic minimization, the Hellmann-Feynman forces in Eqs.~(\ref{eq_frc_KS}), (\ref{eq_frc_efield}) can be evaluated
by invoking the expressions for the total energy in Eq.~(\ref{eq_Eks}) and polarization in Eq.~(\ref{eq_Ptot}), as discussed above.
It should be noted that because the localized orbitals are optimized \emph{in situ} on fixed grids, there is no force contribution
due to the explicit dependence of the basis set on the atomic position --- known as Pulay forces.\cite{ref_Pulay}

According to Eq.~(\ref{eq_Eks}) the KS forces can be separated into two parts, $\vec{F}_{\mathrm{bs}}$ coming from the band-structure
energy term, $E_{\mathrm{bs}}$, and $\vec{F}_{\mathrm{i-i}}$ associated with the ion-ion energy term, $E_{\mathrm{i-i}}$.
Thus, the KS force of the intra- and interatomic interaction acting on the $I$th atom, located at $\vec{\tau}_I$, can be written as
\begin{equation}
 \vec{F}_{\mathrm{KS}}(\vec{\tau}_I) = \vec{F}_{\mathrm{bs}}(\vec{\tau}_I) + \vec{F}_{\mathrm{i-i}}(\vec{\tau}_I) ~.
 \label{eq_frc_KS_bs_ii}
\end{equation}

The force $\vec{F}_{\mathrm{i-i}}(\vec{\tau}_I)$ exerted on one ion by all the other ions can be evaluated as usual in periodic systems by performing
two convergent summations, one over the lattice vectors and the other one over the reciprocal-lattice vectors, using Ewald's method.\cite{ref_frc_Ewald}

If the basis set used to represent the electronic degrees of freedom is independent of the atomic coordinates,
as in our case, $\vec{F}_{\mathrm{bs}}$ is solely due to the explicit dependence of the Hamiltonian on the
position of the atoms. Among the components of the KS Hamiltonian, only the ionic potential, $V_{\mathrm{ion}}$, is a function of $\vec{\tau}$,
thus $\vec{F}_{\mathrm{bs}}=\vec{F}_{\mathrm{ion}}$. This term is evaluated using the pseudopotential theory. We employ nonlocal norm-conserving
ionic pseudopotentials cast in the Kleinman-Bylander form.\cite{ref_KBpseudo} The ionic pseudopotential due to an atom $I$ at position $\vec{\tau}_I$,
$\hat{V}_{\mathrm{ion}}(\vec{\tau}_I)$, is obtained as the sum of a local, $\hat{V}_{\mathrm{loc}}(\vec{\tau}_I)$, and nonlocal, $\hat{V}_{\mathrm{nloc}}(\vec{\tau}_I)$
term, the latter corresponding to an angular-momentum-dependent projection.\cite{ref_KBpseudo}
Like the pseudopotential energy itself, the pseudopotential force is the sum of local and nonlocal parts:
$\vec{F}_{\mathrm{ion}}=\vec{F}_{\mathrm{loc}}+\vec{F}_{\mathrm{nloc}}$.
The local pseudopotential force is given by
\begin{equation}
 \vec{F}_{\mathrm{loc}}(\vec{\tau}_I) = -2\frac{\partial}{\partial\vec{\tau}_I} \sum_{a} \bra{\nu_a} \hat{V}_{\mathrm{loc}}(\vec{\tau}_I) \ket{\tilde\nu_a} ~.
 \label{eq_frc_loc}
\end{equation}

The force component $\vec{F}_{\mathrm{loc}}(\vec{\tau}_I)$ can be calculated as usual by evaluating the expectation
value of the derivative of the ionic potential with respect to the ionic position. \cite{ref_parsec_periodic}
Because $\frac{\partial \hat{V}_{\mathrm{loc}}(\vec{\tau}_I)}{\partial \vec{\tau}_I}$ is local, this reduces to the integration of this term,
multiplied by the electron density, over a grid spanning the simulation cell.

It is however advantageous to consider a different implementation of $\vec{F}_{\mathrm{loc}}$.\cite{ref_Hirose,ref_Octopus} Starting from Eq.~(\ref{eq_frc_loc}),
by changing the integration variable $\vec{r} \rightarrow \vec{r}'=\vec{r}-\vec{\tau}_I$, when evaluating the matrix elements
$\bra{\nu_a} \hat{V}_{\mathrm{loc}}(\vec{\tau}_I) \ket{\tilde\nu_a}$, the $a$th orbital contribution to the local pseudopotential
force on the $I$th ion can be written as
\begin{equation}
 \begin{split}
  & \frac{\partial}{\partial\vec{\tau}_I} \bra{\nu_a} \hat{V}_{\mathrm{loc}}(\vec{\tau}_I) \ket{\tilde\nu_a} = \\
  & \int_{LR_a} \frac{\partial}{\partial\vec{\tau}_I} \nu_a(\vec{r}'+\vec{\tau}_I) V_{\mathrm{loc}}(\vec{r}') \tilde{\nu}_a(\vec{r}'+\vec{\tau}_I) ~\mathrm{d}\vec{r}' ~.
 \end{split}
 \label{eq_frc_loc_int}
\end{equation}
It follows from Eq.~(\ref{eq_frc_loc_int}) that with the transformation of the variable $\vec{r}'$ back to $\vec{r}=\vec{r}'+\vec{\tau}_I$, the local pseudopotential
force can be calculated as
\begin{equation}
 \begin{split}
  \vec{F}_{\mathrm{loc}}(\vec{\tau}_I) = -2 \sum_{a} \bigg[ & \bra{\nabla_{\vec{r}}\nu_a} \hat{V}_{\mathrm{loc}}(\vec{\tau}_I) \ket{\tilde\nu_a} +  \\
                                                            & \bra{\nu_a} \hat{V}_{\mathrm{loc}}(\vec{\tau}_I) \ket{\nabla_{\vec{r}}\tilde\nu_a} \bigg] ~.
 \end{split}
 \label{eq_frc_locr}
\end{equation}
This is the formula for the local pseudopotential force employed in our implementation.

Along the same line as in Eq.~(\ref{eq_frc_locr}), an analogous expression can be derived for the nonlocal pseudopotential force.
The contribution to the force on atom $I$ coming from the nonlocal components of the pseudopotential is
\begin{equation}
 \vec{F}_{\mathrm{nloc}}(\vec{\tau}_I) = -2\frac{\partial}{\partial\vec{\tau}_I} \sum_{a} \sum_{l,m} \braket{\nu_a|\phi_{lm}(\vec{\tau}_I)} \braket{\phi_{lm}(\vec{\tau}_I)|\tilde\nu_a} ~,
 \label{eq_frc_nloc}
\end{equation}
where $\phi_{lm}(\vec{\tau}_I)$ are the projector functions on atom $I$, running over angular momentum indices $l$,~$m$.

By applying the $\vec{r} \rightarrow \vec{r}'=\vec{r}-\vec{\tau}_I$ transformation when evaluating the projection coefficients
$\braket{\phi_{lm}(\vec{\tau}_I)|\nu_a}$ and integrating over the core regions of the $I$th base atom replicas entering the $a$th
orbital localization region, the following formula for the nonlocal pseudopotential force is obtained
\begin{equation}
 \begin{split}
  \vec{F}_{\mathrm{loc}}(\vec{\tau}_I) = -2 \sum_{a} \sum_{l,m} \bigg[ & \braket{\nabla_{\vec{r}}\nu_a|\phi_{lm}(\vec{\tau}_I)} \braket{\phi_{lm}(\vec{\tau}_I)|\tilde\nu_a} +  \\
                                                                       & \braket{\nu_a|\phi_{lm}(\vec{\tau}_I)} \braket{\phi_{lm}(\vec{\tau}_I)|\nabla_{\vec{r}}\tilde\nu_a} \bigg] ~.
 \end{split}
 \label{eq_frc_nlocr}
\end{equation}

Our motivation for using this alternative ionic force formulation, which does not include the derivative of the
ionic potential, but the gradient of the real-space wave functions, as in Eqs.~(\ref{eq_frc_locr}) and (\ref{eq_frc_nlocr}),
is that it is more precise when discretized on a grid. This finding was reported in Ref.~\onlinecite{ref_Octopus},
in the context of molecular calculations employing local ionic potentials. The reason for the improved accuracy can be attributed to
the wave functions that are generally smoother than the ionic potential, so that their derivatives can be better represented on a grid.
The erroneous effect coming from the real-space discretization can be further reduced by employing the pseudopotential filtering technique. \cite{ref_filtering}
It eliminates the Fourier components of the local potential and the projector functions that cannot be represented on the grid.
As an overall result, the convergence of the force calculations with respect to the grid spacing is improved. This is important in view of the
observed moderate convergence of the pseudopotential forces with respect to the LR size of the orbitals, on which the grids are spanned,
and will be further discussed in Sec.~\ref{sec_conv}.

We now turn to the calculation of the force term $\vec{F}_{\mathcal{E}}$ associated with the polarization constraint.
Following the Hellmann-Feynman argument $\vec{F}_{\mathcal{E}}$ can be evaluated according to Eq.~(\ref{eq_frc_efield}),
which involves only the partial derivatives $\frac{\partial\vec{P}}{\partial\vec{\tau}_I}$ of the polarization
$\vec{P}$ in Eq.~(\ref{eq_Ptot}) with respect to the atomic coordinates $\{\vec{\tau}_I\}$.
Since we use a basis set independent of $\vec{\tau}_I$ and employ norm-conserving pseudopotentials
in our calculations, the only explicit dependence of the polarization in Eq.~(\ref{eq_Ptot}) on $\vec{\tau}_I$ comes from the
ionic term $\vec{P}_{\mathrm{ion}}$ given by Eq.~(\ref{eq_Pion}).
This is not true when using ultra-soft pseudopotentials, which need additional augmentation terms in the expression
for the electronic polarization,\cite{ref_pol_uspp} with the explicit dependence on the atomic coordinates.
However, in our case, $\frac{\partial\vec{P}}{\partial\vec{\tau}_I}=\frac{\partial\vec{P}_{\mathrm{ion}}}{\partial\vec{\tau}_I}$,
and consequently the force due to the polarization constraint, acting on atom $I$ is given by
\begin{equation}
 \vec{F}_{\mathcal{E}}(\vec{\tau}_I) = \efield Q_I ~.
 \label{eq_frc_efield_Q}
\end{equation}
As is apparent from Eq.~(\ref{eq_frc_efield_Q}) the $\vec{F}_{\mathcal{E}}(\vec{\tau}_I)$ term is the force induced by
the electric field $\efield$ acting on a point ionic core charge $Q_I$. The same expression is used in finite-field
calculations,\cite{ref_SIV,ref_MD_efield} but here, $\vec{F}_{\mathcal{E}}$ shall be interpreted as a constraint force
with the electric field $\efield$ playing the role of a Lagrange multiplier imposing the polarization constraint.

\section{Solution Algorithm\label{sec_alg}}
In this section an optimization method is proposed that allows to simultaneously relax the atomic coordinates and adjust
the electric field so that the stationary point of the energy functional of the polarization, $E(\vec{P}_t)$, can be found.
The technique is based on the application of the quasi-Newton scheme of Vanderbilt and Louie \cite{ref_VLopt} (VL) to
iteratively solve Eq.~(\ref{eq_frc_pol}). The advantage of the proposed approach is that it does not require the
second derivatives of the enthalpy, which cannot be calculated analytically within DFT. This is in contrast to the
SRV method, which uses DFPT to obtain the guiding tensors in the minimization procedure over structural
degrees of freedom that gives the target polarization.\cite{ref_SRV,ref_DV}

With the VL scheme the roots of a vector function $\vec{f}(\vec{x})$ of a vector variable $\vec{x}$ can be determined.
In the case of Eq.~(\ref{eq_frc_pol}), we define
\begin{equation}
 \vec{x} :=
 \begin{bmatrix}
  \vec{\tau} & \vec{\efield}
 \end{bmatrix}
 ^\top
 \label{eq_xvec_def}
\end{equation}
and
\begin{equation}
 \vec{f} :=
 \begin{bmatrix}
  \vec{F} & \Omega\Delta\vec{P}
 \end{bmatrix}
 ^\top  ~,
 \label{eq_fvec_def}
\end{equation}
where $\Delta\vec{P}=\vec{P}-\vec{P}_t$.

The goal is now to find $\vec{x}$ such that $\vec{f}(\vec{x})=\vec{0}$. For a generally non-linear function
$\vec{f}(\vec{x})$, as in our case, this is done iteratively. Starting from an initial guess
$\vec{x}^{(0)}$ and $\vec{f}^{(0)}=\vec{f}(\vec{x}^{(0)})$, it can be predicted that on the $m$th iteration,
in linear order, a new $\vec{x}^{(m+1)}$ will result in a $\vec{f}^{(m+1)}$ satisfying
the following secant equation \cite{ref_NumRecipes}
\begin{equation}
 \vec{f}^{(m+1)} - \vec{f}^{(m)} = -\mathbf{J} \times \left( \vec{x}^{(m+1)} - \vec{x}^{(m)} \right) ~,
 \label{eq_secant}
\end{equation}
where the Jacobian is defined as
\begin{equation}
 \mathbf{J} = - \frac{\mathrm{d} \vec{f}}{\mathrm{d} \vec{x}} ~.
 \label{eq_jac_def}
\end{equation}
By setting $\vec{f}^{(m+1)}=\vec{0}$, which is the goal of the optimization, gives the Newton step
\begin{equation}
 \vec{x}^{(m+1)} = \vec{x}^{(m)} + [\mathbf{J}^{(m)}]^{-1} \times \vec{f}^{(m)} ~.
 \label{eq_qnstep}
\end{equation}
Eq.~(\ref{eq_qnstep}) can be used to carry out the optimization by iteratively improving the solution vector
$\vec{x}$, if the Jacobian is known.

For the constrained polarization calculation case the Jacobian matrix in Eq.~(\ref{eq_jac_def}) is
composed of the following blocks
\begin{equation}
 \mathbf{J} =
 \begin{bmatrix}
  \mathbf{K}  & -\mathbf{Z} \\
  -\mathbf{Z} & -\Omega \mathbf{\chi}
 \end{bmatrix}
 ~,
 \label{eq_jac_block}
\end{equation}
which correspond to the different components of the $\vec{f}$ and $\vec{x}$ vectors specified in
Eqs.~(\ref{eq_fvec_def}) and (\ref{eq_xvec_def}), respectively.
These blocks have the following interpretation:
$\mathbf{K} := - \frac{\mathrm{d} \vec{F}}{\mathrm{d} \vec{\tau}}$ is the force-constant matrix;
$\mathbf{\chi} := \frac{\mathrm{d} \vec{P}}{\mathrm{d} \efield}$ is the susceptibility tensor;
$\mathbf{Z} := \frac{\mathrm{d} \vec{F}}{\mathrm{d} \efield}$ is the Born effective charge tensor.
Note that the $\mathbf{Z}$ matrix can be re-expressed as
$\mathbf{Z} = - \frac{\mathrm{d}^2 E_{gs}(\vec{\tau},\efield)}{\mathrm{d} \vec{\tau} \mathrm{d} \efield} = \Omega \frac{\mathrm{d} \vec{P}}{\mathrm{d} \vec{\tau}}$
(see Eqs.~(\ref{eq_Egs}) and (\ref{eq_dEgs_HF})), which is used to substitute the second row in Eq.~(\ref{eq_jac_block}).

The matrices $\mathbf{K}$, $\mathbf{\chi}$, and $\mathbf{Z}$ needed to form the Jacobian in Eq.~(\ref{eq_jac_block})
cannot be calculated analytically within DFT because the forces and the polarization are not variational.
Since these matrices are of interest by themselves, for studying the elastic and electric properties of materials,
DFPT techniques were developed to approximate them. These methods consist however of rather involved expressions
which must be carefully handled in the presence of electric fields.\cite{ref_DFPT_efield_I} They further require special
implementations \cite{ref_DFPT_efield_II}. An other common approach to calculate the elements of the force-constant matrix
\cite{ref_phonons}, dielectric, and Born effective charge tensors \cite{ref_btoWf} is by numerical differentiation,
but this is not computationally efficient for the present purposes.

We propose here to use the VL approach and build up the Jacobian through iterative improvements
by employing the information from the previous iterations of the optimization algorithm. Specifically,
the Jacobian $\mathbf{J}^{(m+1)}$ is constructed by requiring that it satisfies in the least-square sense
the secant Eq.~(\ref{eq_secant}) for the $m$ previous iterations. An additional weighted condition
stating that $\mathbf{J}^{(m+1)}$ makes the least change to the initial Jacobian $\mathbf{J}^{(0)}$
is also imposed
\begin{equation}
 \begin{split}
  S = & \sum_{l=0}^m w^{(l)} \left| \mathbf{J}^{(m+1)}\times\Delta\vec{x}^{(l)} + \Delta\vec{f}^{(l)} \right|^2 \\
      & + w^{(0)} \left\| \mathbf{J}^{(m+1)}-\mathbf{J}^{(0)} \right\|^2 ~,
 \end{split}
 \label{eq_jac_ls}
\end{equation}
where
$\Delta\vec{x}^{(l)} = \left( \vec{x}^{(l+1)} - \vec{x}^{(l)} \right) / \left| \vec{x}^{(l+1)} - \vec{x}^{(l)} \right|$
and
$\Delta\vec{f}^{(l)} = \left( \vec{f}^{(l+1)} - \vec{f}^{(l)} \right) / \left| \vec{f}^{(l+1)} - \vec{f}^{(l)} \right|$
represent the normalized differences of the successive iterations.
We use $\left|\vec{x}\right|$ to denote the $L^2$-norm of a vector $\vec{x}$ ($\left|\vec{x}\right|=\sqrt{\sum_i x_i^2}$),
and $\left\|\mathbf{M}\right\|$ to denote the Frobenius norm of a matrix $\mathbf{M}$ ($\left\|\mathbf{M}\right\|=\sqrt{\sum_{ij}M_{ij}^2}$).

The least-squares minimization problem in Eq.~(\ref{eq_jac_ls}) can be solved for the updated Jacobian $\mathbf{J}^{(m+1)}$
by setting ${\partial S}/{\partial J_{ij}^{(m+1)}}=0$, which gives
\begin{equation}
 \mathbf{J}^{(m+1)} = \mathbf{A}^{(m+1)} \times [\mathbf{B}^{(m+1)}]^{-1} ~,
 \label{eq_jac_step}
\end{equation}
where
\begin{alignat*}{2}
 \mathbf{A}^{(m+1)} &= w^{(0)}\mathbf{J}^{(0)} & &- \sum_{l=0}^{m} \Delta\vec{f}^{(l)} \otimes [\Delta\vec{x}^{(l)}]^\top ~, \\
 \mathbf{B}^{(m+1)} &= w^{(0)}\mathbf{I} & &+ \sum_{l=0}^{m} \Delta\vec{x}^{(l)} \otimes [\Delta\vec{x}^{(l)}]^\top ~.
\end{alignat*}
In the above equations $\mathbf{I}$ denote the identity matrix.

In the limit where only the most recent iteration is used to update the Jacobian ($l=m$ in Eq.~(\ref{eq_jac_ls}))
and $w^{(0)} \ll 1$, the VL method reduces to the Broyden-Fletcher-Goldfarb-Shanno (BFGS) Jacobian updating scheme.\cite{ref_ModifiedBFGS}
We favor the VL method for its stability and efficiency, which will be illustrated in Sec.~\ref{sec_conv}.

By combining Eqs.~(\ref{eq_qnstep}) and (\ref{eq_jac_step}) the constrained polarization calculation
can be carried out, starting from a trial guess $\vec{\tau}^{(0)}$, $\efield^{(0)}$, and $\mathbf{J}^{(0)}$.
At each step $m$ of the algorithm, the electronic degrees of freedom are optimized by the minimization of the
electric enthalpy in Eq.~(\ref{eq_Egs}) at current atomic configuration $\vec{\tau}^{(m)}$ and electric field $\efield^{(m)}$.
Subsequently, the Hellmann-Feynman forces $\vec{F}^{(m)}$ and polarization $\vec{P}^{(m)}$ are calculated using
Eqs.~(\ref{eq_frc_tot}) and (\ref{eq_Ptot}), respectively.
With them Eq.~(\ref{eq_qnstep}) can be formed, which gives the new $\vec{\tau}^{(m+1)}$ and $\efield^{(m+1)}$.
Then $\vec{F}^{(m+1)}$ and $\vec{P}^{(m+1)}$ are re-evaluated at $(\vec{\tau}^{(m+1)} , \efield^{(m+1)})$,
and the Jacobian is updated according to Eq.~(\ref{eq_jac_step}). The result, $\mathbf{J}^{(m+1)}$, is used to solve
Eq.~(\ref{eq_qnstep}) in the next iteration of the algorithm, $m\leftarrow{}m+1$. This refining procedure of
$\vec{\tau}^{(m)}$ and $\efield^{(m)}$ continues until $\vec{F}^{(m)}$ and $\Delta\vec{P}^{(m)}$ both vanish.
After convergence is reached, the algorithm moves to the next polarization constraint.

The Jacobian updating scheme in Eq.~(\ref{eq_jac_step}) requires an initial guess $\mathbf{J}^{(0)}$.
It has to be set properly to assure reasonable step size during the first few optimization iterations.
We construct $\mathbf{J}^{(0)}$ from a diagonal force-constant matrix $\mathbf{K}^{(0)}=k^{(0)}\mathbf{I}$,
diagonal susceptibility tensor $\mathbf{\chi}^{(0)}=\chi^{(0)}\mathbf{I}$, and Born effective charges equal
to guessed static atomic charges $\mathbf{Z}^{(0)}$. The values of the $k^{(0)}$ and $\chi^{(0)}$
diagonal scaling factors and the elements of the $\mathbf{Z}^{(0)}$ matrix are the only free parameters
of the algorithm, which makes it straightforward to use. There are certainly more sophisticated ways of initializing
$\mathbf{J}^{(0)}$, for instance by using surrogate models,\cite{ref_preconditioners} e.g. force fields \cite{ref_ModelHessian}
combined with the polarizability model of Bilz et al. \cite{ref_PolarizabilityModel} for the present purposes.
However, this only matters during the first few relaxation steps of a new structure. Moreover, if the calculation
is to be done for multiple polarization target values, as is usually the case, the fully buildup $\mathbf{J}$-matrix
can be passed from one calculation to the next and used as an initial guess to further improve the performance,
as will be shown in Sec.~\ref{sec_conv}.

\section{Computational details\label{sec_comput}}

The method described in the previous sections has been implemented in our in-house version \cite{ref_my_NGWF} of
the \texttt{\MakeUppercase{parsec}} open-source DFT code.\cite{ref_PARSEC_web}
We have applied it to study the ferroelectric properties of \BTO{} and \PTO{}, as outlined below.

\begin{figure}[h]
\centering
\begin{tikzpicture}
\node[anchor=south west,inner sep=0] (image) at (0,0) {\includegraphics[width=0.5\linewidth]{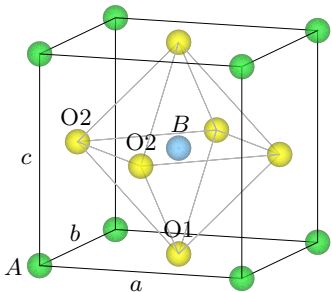}};
\begin{scope}[x={(image.south east)},y={(image.north west)}]
 \node[below] at (0.37,0.095) {$a$};
 \node[left] at (0.075,0.465) {$c$};
 \node[above left] at (0.225,0.17) {$b$};
 \node at (-0.007,0.112) {$A$};
 \node at (0.505,0.580) {$B$};
 \node at (0.50,0.235) {$\mathrm{O1}$};
 \node at (0.185,0.60) {$\mathrm{O2}$};
 \node at (0.385,0.52) {$\mathrm{O2}$};
\end{scope}
\end{tikzpicture}
\caption{Tetragonal unit cell of $AB\mathrm{O_3}$ perovskite in the unsymmetrical phase.
Atoms $A$, $B$, and $\mathrm{O}$ are represented by green, light blue, and yellow spheres, respectively.
The symbols $a$, $b$, and $c$ are the lattice constants.
The labels next to the atoms indicate the positions of the atomic sites used to calculate the polarization quantum.
The fractional coordinates, $(\frac{x}{a},\frac{y}{b},\frac{z}{c})$, of these locations are
$(0,0,0)$ for $A$,
$(0.5,0.5,0.5)$ for $B$,
$(0.5,0.5,0.0)$ for $\mathrm{O1}$,
$(0.0,0.5,0.5)$ and $(0.5,0.0,0.5)$ for two equivalent $\mathrm{O2}$ sites.
\label{fig_struct}}
\end{figure}

The unit cell of the considered perovskite compounds, with the formula $AB\mathrm{O_3}$, is depicted in Fig.~\ref{fig_struct}.
A tetragonal structure is assumed for both \PTO{} and \BTO{}, which corresponds to their room temperature phase.
The experimental lattice parameters used in this study are listed in Table~\ref{tab_struct}.
The lattice is kept fixed during the calculations.

\begin{table}[h]
 \centering
 \caption{Experimental lattice parameters of the tetragonal perovskite structure of \BTO{} \cite{ref_BTO_exp_struct} and \PTO{} \cite{ref_PTO_exp_struct}.}
 \begin{ruledtabular}
 \begin{tabular*}{\linewidth}{@{\extracolsep{\fill} } cccc}
  $AB\mathrm{O_3}$ & $a$, $b$ $[\mathrm{r_B}]$ & $c/a$ & Volume $[\mathrm{r_B^3}]$ \\
  \hline
  \BTO{}             &  7.53  & 1.01 & 431.69 \\
  \PTO{}             &  7.38  & 1.06 & 427.08 \\
 \end{tabular*}
 \end{ruledtabular}
\label{tab_struct}
\end{table}

The atomic positions depicted in Fig.~\ref{fig_struct} are those of the centrosymmetric, non-polar reference
state of $AB\mathrm{O_3}$ perovskite oxides. The absolute values of the polarization components along the lattice vectors
calculated for this paraelectric state are used to define the polarization quanta,\cite{ref_MTP_guide} $P_i^q$, for a chosen unit cell.
The actual values of the polarization, as reported in Sec.~\ref{sec_results}, are calculated by subtracting $P_i^q$ from the $P_i$ values
computed using Eq.~(\ref{eq_Ptot}), along the corresponding lattice vectors.
This treatment allows to remove the ``modulo $1/\Omega$ times real-space lattice vector'' ambiguity \cite{ref_MTP_piezo} in the expression
for $\vec{P}$ in Eq.~(\ref{eq_Ptot}). The polarization calculated in this way is well-defined and independent of the choice of the
unit cell. Note that this redefinition of $\vec{P}$ agrees with the modern theory of polarization,
which considers only the changes of the polarization with respect to a reference state.\cite{ref_MTP_KSV_I,ref_MTP_Resta,ref_MTP_KSV_II}

In this work the pseudopotential model of a solid \cite{ref_pp_rev} is employed to describe the constituent atomic species.
We utilize nonlocal norm-conserving ionic pseudopotentials cast in the real-space Kleinman-Bylander form,\cite{ref_KBpseudo}
to evaluate the interactions between the ion cores and the valence electrons.
The pseudopotentials are generated with the \texttt{atom} software \cite{ref_atom_web} and rely on the Troullier-Martins prescription.\cite{ref_TMpseudo}
Since this code only allows for one pseudized state per angular momentum channel, the semicore states of the
$\mathrm{Ba}$, $\mathrm{Pb}$, and $\mathrm{Ti}$ atoms are kept in the core. Partial core correction \cite{ref_CoreCorrect}
is included for these atoms in order to improve the quality of the solid-state calculations.
The force term resulting from the use of a partial core correction is included into the calculations of the total forces acting on the
$\mathrm{Ba}$, $\mathrm{Pb}$, and $\mathrm{Ti}$ atoms. This step is necessary due to the introduction of an
explicit dependence of the exchange-correlation functional on atomic positions via the atom-centered core
charge densities.\cite{ref_PARSEC_polar}

In the frozen-core approximation that underlies the pseudopotential theory, the species placed at the atomic sites
are the effective ions. They are composed of the nucleus and the electron cores. For the chosen configurations of the
pseudopotentials, the net positive charge of the nucleus plus core is:
$+2$ for $\mathrm{Ba}$,
$+4$ for $\mathrm{Pb}$,
$+4$ for $\mathrm{Ti}$, and
$+6$ for $\mathrm{O}$.
These values of the atomic valence charges are used when evaluating the ionic polarization in Eq.~(\ref{eq_Pion}) and
forces due to the electric field in Eq.~(\ref{eq_frc_efield_Q}).

Within the pseudopotential approximation to DFT only the valence electrons are explicitly accounted for in the
solid-state calculations. For the chosen valence-core partition the number of valence electrons per unit cell is
$N=24$ in the case of \BTO{} and $N=26$ for \PTO{}. In our implementation \cite{ref_my_NGWF} the valence electrons are represented by
Wannier-like orbitals, which are calculated on uniform real-space grids spanning the localization regions (LRs).
The wave functions are truncated beyond the LRs boundaries. This is the only additional approximation as compared to
standard real-space pseudopotential DFT calculations.\cite{ref_PARSEC_FD}
The LRs have a tetragonal shape and their size is determined by the multiple of unit cells, $N_{\mathrm{cell}}$,
that enter the LRs. The positions of the LRs are set at the beginning of the simulation and are kept fixed
during the whole process.
The number of LRs is equal to the number of occupied orbitals. In the present work a double-occupancy of the orbitals
is assumed. This implies that the $N=24$ valence electrons in \BTO{} and $N=26$ valence electrons in \PTO{}
unit cells are covered by $\frac{N}{2}=12$ and $\frac{N}{2}=13$ orbitals, respectively.
In both compounds the LRs for $12$ of these orbitals are centered on the $\mathrm{O1}$ and $\mathrm{O2}$ sites shown
in Fig.~\ref{fig_struct}. The $4$ orbitals per $\mathrm{O}$ atom site are initialized with Gaussians having a $s$, $p_x$, $p_y$, and $p_z$
symmetry, and an origin on the central $\mathrm{O}$ atom. The additional orbital in \PTO{} is calculated in a LR centered
on the $\mathrm{Pb}$ atom located at the $A$ site in Fig.~\ref{fig_struct}. The initial guess for this orbital is taken to be a
spherically symmetric Gaussian function.

Before proceeding, we should acknowledge an additional theoretical subtlety associated with the correct choice of the
exchange-correlation functional in the electric-field problem. The currently accepted view \cite{ref_DFPT_response,ref_DFPT_extended,ref_DFPT_exc}
is that a dependence on the polarization should be present in the exchange-correlation functional --- leading to a
density-polarization functional theory. However, no realistic polarization-dependent functional has been proposed
yet and in this work we remain at the standard LDA level. We use the exchange-correlation functional of
Ceperley and Alder,\cite{ref_LDA_CA} as parameterized by Perdew and Zunger.\cite{ref_LDA_PZ}

\section{Results\label{sec_results}}

In this section, the computational scheme outlined above is illustrated by applying it to a series of problems
involving constrained polarization calculations in tetragonal perovskite compounds.
First, the accuracy of the calculations is verified and the performance of the optimization algorithm is examined.
Then, results concerned with the ferroelectric behavior of \BTO{} and \PTO{} are presented and the differences in
the ferroelectric properties between the materials are studied.

\subsection{Numerical tests\label{sec_conv}}

The reliability of structural relaxation calculations is determined by the accuracy of the atomic forces.
Directly solving for localized orbitals by constraining the wave functions to be zero outside the localization regions (LRs)
introduces an error in the total energy,\cite{ref_KMG} which is expected to be present in the case of forces too.
It has been shown by the authors in Ref.~\onlinecite{ref_my_NGWF} that the errors due to the localization constraint can be alleviated
by allowing the localized wave functions to be non-orthogonal, what leads to non-orthogonal generalized
Wannier functions (NGWFs). In particular, it has been demonstrated that total energy calculations in
the presence of electric field converge faster with increasing LRs size when using NGWFs instead of orthogonal
Wannier functions (WFs).

\begin{figure}[h]
\centering
\includegraphics[width=0.99\linewidth]{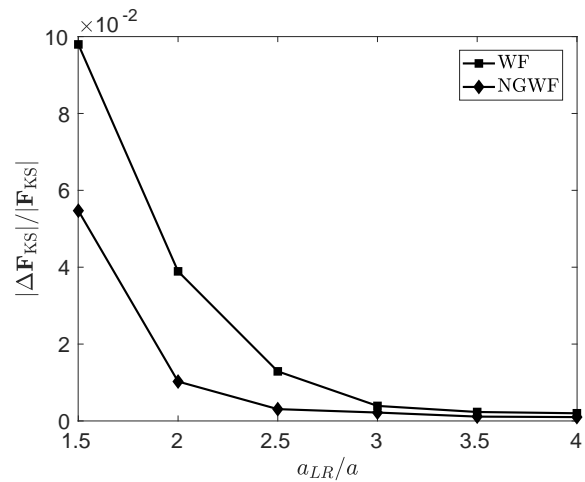}
\caption{Convergence of atomic forces as a function of the LR size in tetragonal \BTO{} at a fixed, off-equilibrium geometry.
The forces are induced by displacing the $\mathrm{Ti}$ atom in the centrosymmetric structure of Fig.~\ref{fig_struct} by $\Delta_z=0.1\times{} c$.
$\vec{F}$ is the force vector calculated with localized orbitals, either NGWFs (line with diamonds) or WFs (line with squares). The reference forces
$\vec{F^{\mathrm{ref}}}$ were obtained using Bloch functions.
\label{fig_BTO_frc}}
\end{figure}

The study carried out in Ref.~\onlinecite{ref_my_NGWF} considered clamped atoms in their equilibrium structure.
In Fig.~\ref{fig_BTO_frc} we show how the localization error affects the total forces induced by the atomic distortion.
The off-equilibrium structure results from a displacement of the $\mathrm{Ti}$ atom from its centrosymmetric position
in Fig.~\ref{fig_struct} by $\Delta_z=0.1\times{} c$.
The quantity plotted in Fig.~\ref{fig_BTO_frc} is the relative deviation of the forces computed with NGWFs and WFs
from the values obtained using Bloch functions, which are taken to be a converged result.
In this case Brillouin zone integrations are performed on a $3\times3\times3$ Monkhorst-Pack mesh \cite{ref_MPgrid}
and the pseudopotential force terms are calculated in a standard way, from the derivatives of the ionic potentials \cite{ref_parsec_periodic}.
We note that thanks to the improved numerical accuracy of the alternative force formulas in Eqs.~(\ref{eq_frc_locr})~and~(\ref{eq_frc_nlocr}),
when discretized on the real-space grid, a coarser grid can be used to obtain the forces appropriately converged with respect to the grid spacing.
The grid step required to converge the forces is $0.3a$ and $0.15a$ when using the alternative and standard scheme to calculate the pseudopotential
forces, respectively. Hence, the number of required grid points decreases by a factor of $8$ to represent each wave function.

Figure~\ref{fig_BTO_frc} shows that the error in forces goes to zero with increasing LRs size, as expected. We also see that the error
decays much faster for NGWFs than WFs. The convergence of forces with respect to the LR size is somewhat slower
than that of the total energy and polarization. It thus determines the overall accuracy of the calculations.
As can be seen in Fig.~\ref{fig_BTO_frc}, the forces obtained from NGWFs are already in good agreement with the reference result,
with no localization constraint, setting $a_{\mathrm{LR}}=2.5a$. In this case the relative deviation is less than $0.3\%$.
For the forces calculated with WFs $a_{\mathrm{LR}}=3.5a$ is required to reach a similar accuracy. This amounts to a
volume difference by a factor of $2.7$ to represent each wave function.

\begin{figure}[h]
\centering
\includegraphics[width=0.99\linewidth]{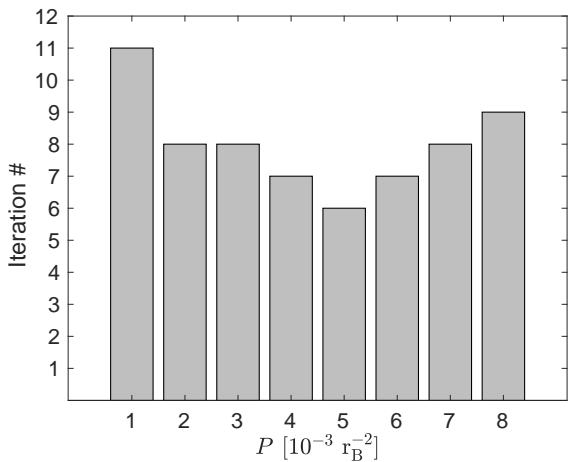}
\caption{Number of iterations required to converge the quasi-Newton algorithm for constrained polarization
calculations in tetragonal \BTO{}.
Starting from the centrosymmetric perovskite structure under zero electric field conditions calculations at consecutive
polarization constraints $P_t$ are performed.
The solution variables and the Jacobian are carried over from one calculation to the next.
In the first one the $\mathbf{J}^{(0)}$ matrix from Sec.~\ref{sec_alg} is used to initialize the Jacobian.
\label{fig_BTO_iter}}
\end{figure}

The efficiency of our method for performing constrained polarization calculations in the case of tetragonal \BTO{}
is illustrated in Fig.~\ref{fig_BTO_iter}. The starting configuration for the subsequent simulations is the centrosymmetric
perovskite structure in Fig.~\ref{fig_struct}, under zero electric field. The required polarization constraints $P_t$
are aligned with the $c$ axis. The initial Jacobian $\mathbf{J}^{(0)}$ is set according to the prescription
given in Sec.~\ref{sec_alg}, with a diagonal force constant matrix formed from $k^{(0)}=0.1$, a diagonal susceptibility tensor
scaled by $\chi^{(0)}=5.0$, and Born effective charges substituted by nominal valences
$+2$ for $\mathrm{Ba}$,
$+2$ for $\mathrm{Pb}$,
$+4$ for $\mathrm{Ti}$, and
$-2$ for $\mathrm{O}$.
The weight associated with $\mathbf{J}^{(0)}$ is $w^{(0)}=0.01$ and the information from all previous iterations
is incorporated when evaluating the Jacobian update according to Eq.~(\ref{eq_jac_step}).
The $\mathbf{J}$-matrix is carried over from one converged relaxation to the next.
The convergence criterion is such that the forces are smaller than $10^{-3}\mathrm{Ry/r_B}$,
and the polarization fulfills the constraint by at least $10^{-5}\mathrm{r_B^{-2}}$.

There are six degrees of freedom to optimize in the calculations. They consist of the internal $z$ coordinates
of five atoms in the perovskite unit cell and the electric field component along the $c$ axis.
It is thus a good example to show the performance of our method, because there are enough degrees of freedom to
make a direct minimization impractical.
As can be seen in Fig.~\ref{fig_BTO_iter} the convergence of our method is achieved after $6$ to $11$ iterations for all
considered polarization constraints, which, as will be shown in Sec.~\ref{sec_BTO}, are sufficient to extract the energy
landscape of tetragonal \BTO{}. The increased number of iterations in the first configuration, at $P_t=1\times10^{-3}~\mathrm{r}_{B}^{-2}$,
as compared to the following ones, is due to the fact that the $\mathbf{J}^{(0)}$ matrix is used to
start the iteration. In the next cases a better initial guess for the Jacobian can be employed,
taken from a fully built-up $\mathbf{J}$-matrix resulting from the previous polarization constraint. Another noticeable
feature of the convergence behavior of the algorithm is the decreased number of iterations in the vicinity
of $P_t=5\times10^{-3}~\mathrm{r}_{B}^{-2}$: this polarization constraint is close to the minimum
of the potential energy well. As a consequence, the harmonic approximation leading to the quasi-Newton step in Eq.~(\ref{eq_qnstep})
is best fulfilled around this point.

\begin{figure}[h]
\centering
\includegraphics[width=0.99\linewidth]{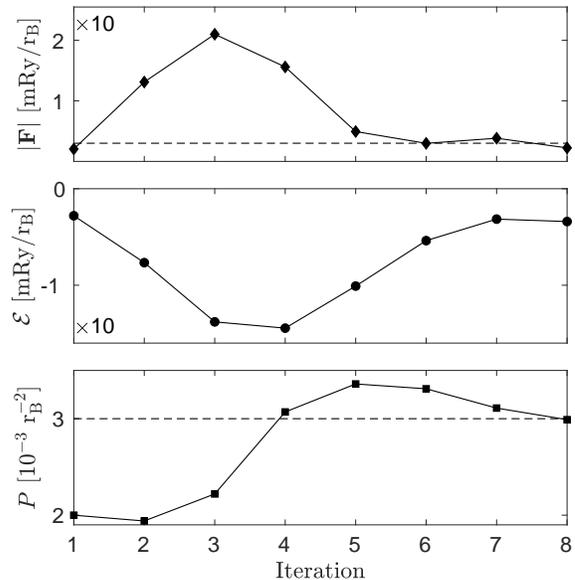}
\caption{Relaxation of tetragonal \BTO{} at $P_t=3\times10^{-3}~\mathrm{r}_{B}^{-2}$. The starting configuration and
the initial Jacobian are taken from a converged calculation at $P_t=2\times10^{-3}~\mathrm{r}_{B}^{-2}$.
Top~panel: \mbox{$L^2$-norm} of the forces acting on the atoms. Middle~panel: electric field component along the $c$ axis.
Bottom~panel: polarization along $\efield$. The convergence threshold for the forces and the polarization constraint
are indicated by the dashed lines in the top and bottom panels, respectively.
\label{fig_BTO_conv}}
\end{figure}

As an example of the algorithm functionality, Fig.~\ref{fig_BTO_conv} shows the convergence of the $L^2$-norm of all calculated forces,
together with the evolution of the electric field and polarization as a function of the number of iterations.
This numerical experiment corresponds to going from a polarization constraint of $P_t=2\times10^{-3}~\mathrm{r}_{B}^{-2}$ to
$P_t=3\times10^{-3}~\mathrm{r}_{B}^{-2}$. During the first iterations the electric field increases in magnitude due to the difference between
the polarization value and the constraint. Note the correct direction of change of the electric field and the polarization from the first iteration.
This feature can be attributed to the correct information about the electric enthalpy surface contained in the $\mathbf{J}$-matrix carried over
from the previous calculation. The change in the electric field induces the forces
\pagebreak
responsible for moving the~atoms.
The relaxation proceeds by displacing the atoms and adjusting the electric field so that the atomic forces
are balanced and the polarization constraint is simultaneously satisfied.

\subsection{Ferroelectricity in tetragonal $\mathbf{BaTiO_3}$\label{sec_BTO}}

\begin{figure}[!htbp]
\centering
\includegraphics[width=0.99\linewidth]{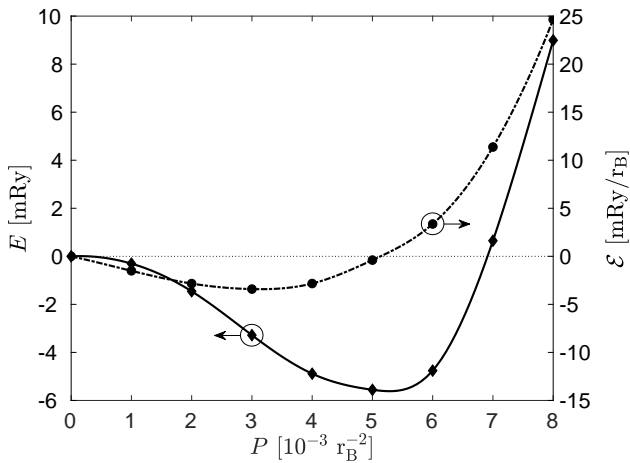}
\caption{Calculated energy (diamonds) and electric-field (circles) as a function of the polarization in tetragonal \BTO{}.
The energy $E$ is the Kohn-Sham total energy in Eq.~(\ref{eq_Eks}) per $5$-atom unit cell.
The electric field $\mathcal{E}$ is aligned with the $c$ axis and imposes the polarization
argument $P$ along the same axis.
\label{fig_BTO_Efield}}
\end{figure}

Having verified the convergence of our method, we now demonstrate its utility by analyzing the energy landscape of
tetragonal \BTO{}. Figure~\ref{fig_BTO_Efield} shows the calculated cross section of the potential energy surface as a function of
the polarization along the $c$ axis. The electric field component along the same axis, which is required to realize the polarization
states, is also plotted in Fig.~\ref{fig_BTO_Efield}. As can be seen, the minimum of the energy is at non-zero value of $P$.
This is the typical signature of ferroelectricity. This minimum coincides with the zero-crossing of $\mathcal{E}$ and corresponds to the
equilibrium state of the material. The value of the spontaneous polarization at this point, $P_s=5\times10^{-3}~\mathrm{r_B}^{-2}$,
in SI units, can be converted to $0.286~\mathrm{C}/\mathrm{m}^2$ and agrees well with the experimental value \cite{ref_BTO_exp_pol} of
$0.26~\mathrm{C}/\mathrm{m}^2$. For $P>P_s$ the state can be realized by applying an appropriate fixed electric field $\mathcal{E}>0$.
The states with $P<P_s$ and $\mathcal{E}<0$ are local maxima of the electric enthalpy and thus cannot be reached by a direct application
of the electric field. It has been recently proposed that ferroelectric materials can be biased into the $\left\{P<P_s,\mathcal{E}<0\right\}$
state by putting them in series with a dielectric \cite{ref_NegCap}. In this case $\mathcal{E}$ acts as a depolarizing field due to the
incomplete screening of the ferroelectric polarization.

In addition to the physical consistency of our results we also report a good agreement between our calculated potential energy
curve and the one obtained by DV in Ref.~\onlinecite{ref_DV}. DV found the energy minimum to be located at
${P_s=(5\pm0.1)\times10^{-3}\mathrm{r_B}^{-2}}$ with a well depth ${E_{\mathrm{min}}=(6\pm0.1)\mathrm{mRy}}$.
Similarly, in our calculations the absolute value of the energy minimum at ${P_s=5\times10^{-3}\mathrm{r_B}^{-2}}$
is ${E_{\mathrm{min}}=5.56~\mathrm{mRy}}$. Given the difference in the pseudopotentials used and our choice of the lattice parameter,
it cannot be excluded that, this level of agreement is partly fortuitous as the ferroelectric potential energy surfaces
are quite sensitive to the choice of the lattice constant and to the approximations in first-principles calculations.
Furthermore, different methods are used in in Ref.~\onlinecite{ref_DV} and here.

\begin{figure}[!htbp]
\centering
\includegraphics[width=0.99\linewidth]{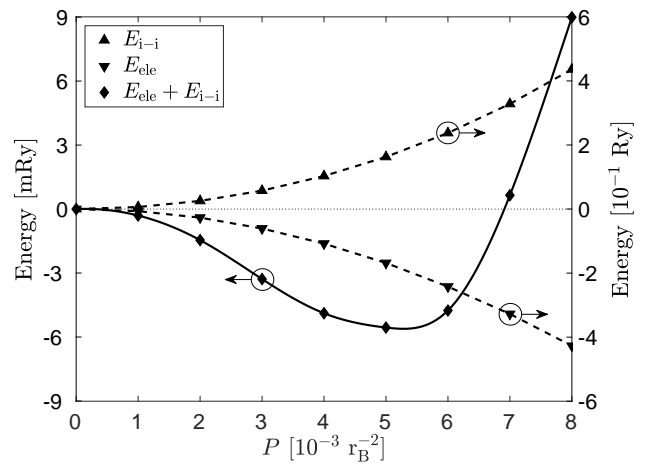}
\caption{Decomposition of the total energy in \BTO{} into the electronic and ion-ion terms.
The electronic energy $E_{\mathrm{ele}}$ is the band-structure term minus double-count corrections.
The component $E_{\mathrm{i-i}}$ is the Ewald energy among the ion cores.
The sum $E_{\mathrm{ele}}+E_{\mathrm{i-i}}$ is the total Kohn-Sham energy in Eq.~(\ref{eq_Eks}).
\label{fig_BTO_Edecomp}}
\end{figure}

In order to make a better understanding of the \BTO{} energy landscape, in Fig.~\ref{fig_BTO_Edecomp} we have decomposed the total energy
into its electronic and ion-ion interaction contributions. It is notable from this figure that the ion-ion repulsion
energy increases with $P$, whereas the electronic energy decreases.
The lowering of the electronic energy can be related to the process of covalent bond formation (hybridization),
what will be shown later in this section.
The total energy is the sum of both contributions. As can be deduced from Fig.~\ref{fig_BTO_Edecomp},
in order to produce a ferroelectric potential well, the electronic energy must decrease faster than the ion-ion
energy increases for $P<P_s$. Thus, if the hybridization processes are not strong enough the ferroelectric state
is not stabilized. For $P>P_s$ the short-range repulsions start to dominate and consequently the total
energy increases with $P$.

\begin{figure}[!htbp]
\centering
\includegraphics[width=0.99\linewidth]{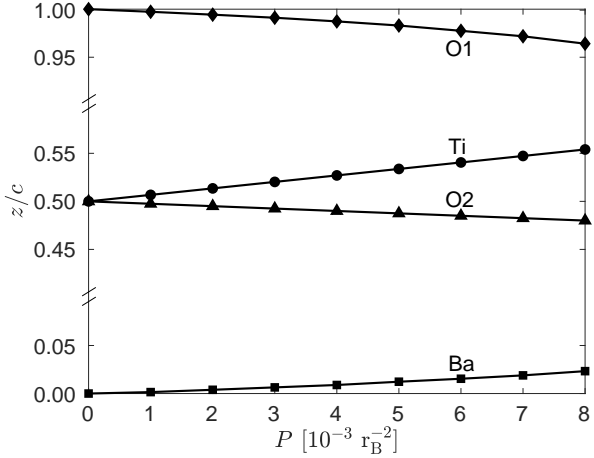}
\caption{Internal $z$ coordinates for each atom in \BTO{} unit cell as a function of polarization $P$ along the $c$ axis.
For initial atomic positions, at $P=0$, see Fig.~\ref{fig_struct}. Top-most points (diamonds) correspond to translational
image of O1 atom in neighboring unit cell. The coordinates at $P=5\times10^{-3}~\mathrm{r_B}^{-2}$
give the equilibrium positions of the atoms.
\label{fig_BTO_zat}}
\end{figure}

The variation of the total energy is due to atomic displacements and the changes in the electronic structure.
Figure~\ref{fig_BTO_zat} shows the polarization dependence of the atomic coordinates in \BTO{} for tetragonal distortions
along the $c$ axis. The atomic displacement pattern displayed in Fig.~\ref{fig_BTO_zat} is such that the $\mathrm{O1}$ and $\mathrm{O2}$
atoms move downward, in a direction opposite to the $\mathrm{Ba}$ and $\mathrm{Ti}$ atoms, which move upward
with increasing $P$. For $P \leq P_s$ the atoms listed in order of increasing displacements from the initial
positions are \mbox{$\{ \mathrm{Ba}$, $\mathrm{O2}$, $\mathrm{O1}$, $\mathrm{Ti} \}$}. This sequence changes for $P > P_s$
as the magnitude of the $\mathrm{Ba}$ displacement exceeds that of $\mathrm{O2}$, changing the character of the
structural distortion. At $P=P_s$ the fractional displacements of $\mathrm{Ti}$, $\mathrm{O1}$, and $\mathrm{O2}$ atoms
with respect to $\mathrm{Ba}$
\mbox{$(\frac{\Delta_{z}(\mathrm{Ti})}{c},\frac{\Delta_{z}(\mathrm{O1})}{c},\frac{\Delta_{z}(\mathrm{O2})}{c})_{\mathrm{Ba}}$}
are found to be $(0.019,-0.029,-0.023)$.
The calculated values in the equilibrium state are in good overall agreement with the experimentally measured displacements \cite{ref_BTO_exp_struct}
$(0.015,-0.023,-0.014)$. As can be seen, the displacement pattern is preserved between the calculated and experimental
tetragonal distortions. This finding is consistent with the results of standard structural relaxations within
DFT and LDA at experimental lattice parameters, which predict equilibrium displacements \cite{ref_ferro_dft_rev}
$(0.013,-0.025,-0.016)$, in agreement with our results at $P=P_s$.

\begin{figure}[!htbp]
\centering
\includegraphics[width=0.99\linewidth]{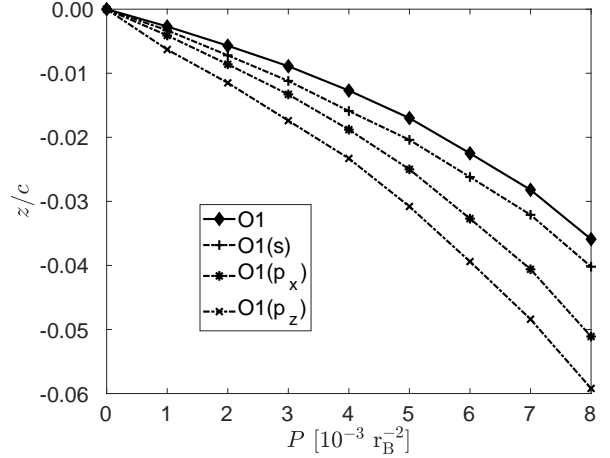}
\caption{Computed $z$ coordinate of the $\mathrm{O1}$ orbitals centroids of charge (dashed lines) and of
$\mathrm{O1}$ atom (solid line) in \BTO{} as a function of the polarization $P$ along the $c$ axis.
The orbitals are labeled by their dominant atomic character on the $\mathrm{O1}$~atom.
The coordinates of $\mathrm{O1}(p_y)$ OC overlap with those of $\mathrm{O1}(p_x)$~OC.
\label{fig_BTO_zwf}}
\end{figure}

The atomic movement induce the electronic charge redistribution. This results in the displacement of the
centroids of charge of the orbitals (OCs). As for the orbitals initially centered on the $\mathrm{O2}$ atomic sites,
we observe that the OCs of $4$ orbitals per each $\mathrm{O2}$ atom follow it when moving. On average, a charge of $-8$ electrons
(due to the double occupancy of the orbitals) moves together with the $+6$ point charge of the $\mathrm{O}$ ion so that it can be
effectively treated as an anion carrying $-2$ charge. Hence, when the $\mathrm{O2}$ atom is displaced in the negative direction
along the coordinate axis, as in Fig.~\ref{fig_BTO_zat}, it gives rise to a positive contribution to the total polarization.

\begin{figure}[!htbp]
\subfloat[paraelectric\label{fig_BTO_pz_para}]{%
  \includegraphics[width=0.22\textwidth,clip=true]{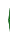}
}
\hfill
\subfloat[ferroelectric\label{fig_BTO_pz_ferro}]{%
  \includegraphics[width=0.22\textwidth,clip=true]{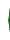}
}
\colorcaption{Amplitude isosurface plots of the $\mathrm{O1}(p_z)$ WFs in \BTO{} at $\pm0.015~\mathrm{r_B}^{-3/2}$.
Red and blue surfaces correspond to positive and negative amplitudes, respectively.
The orbitals are oriented along the $\mathrm{O}$--$\mathrm{Ti}$--$\mathrm{O}$--$\mathrm{Ti}$--$\mathrm{O}$ chains.
$\mathrm{O}$ is at the center, embedded in the $p_{z}$ atomic orbital; above and below are the $\mathrm{Ti}$ atoms (light blue),
almost hidden under the $d_{z^2}$ orbitals; the two other $\mathrm{O}$ atoms (yellow) are located at the top and bottom.
The four $\mathrm{Ba}$ atoms (light green) neighboring the central oxygen are also shown.
(a) Paraelectric state at $P=0$. (b) Ferroelectric state at $P=P_s$.
\label{fig_BTO_pz}}
\end{figure}

The situation is more interesting for the $\mathrm{O1}$ orbitals, centered initially on the $\mathrm{O1}$ atomic site.
The displacements along the $c$ axis of the computed OCs for these orbitals are plotted as a function of $P$ in Fig.~\ref{fig_BTO_zwf}.
The labeling of the orbitals is according to the dominant atomic character on the $\mathrm{O1}$ site in the centrosymmetric
structure. The displacement of the $\mathrm{O1}$ atom is also shown for reference. As it can be observed, when the $\mathrm{O1}$
atom moves in the $-z$ direction, the OCs of the $\mathrm{O1}$ orbitals are displaced downward, even more than the $\mathrm{O1}$
atom has moved. The shift of the $\mathrm{O1}$ OCs is towards $\mathrm{Ti}$ atom, which moves upward, in the direction of
the $\mathrm{O1}$ atom (see Fig.~\ref{fig_BTO_zat}). This relative displacement of the $\mathrm{O1}$ OCs with respect to the moving atoms
leads to a larger positive contribution to the total polarization than if the $\mathrm{O1}$ OCs would move rigidly with the
$\mathrm{O1}$ ion. To investigate the physical processes modulating the amplitude of the $\mathrm{O1}$ OCs displacement,
which enhance the electronic polarization, we will now visualize the $\mathrm{O1}$ orbitals and study their transformations
induced by the atomic displacement.

\begin{figure}[!htbp]
\centering
\includegraphics[width=0.99\linewidth]{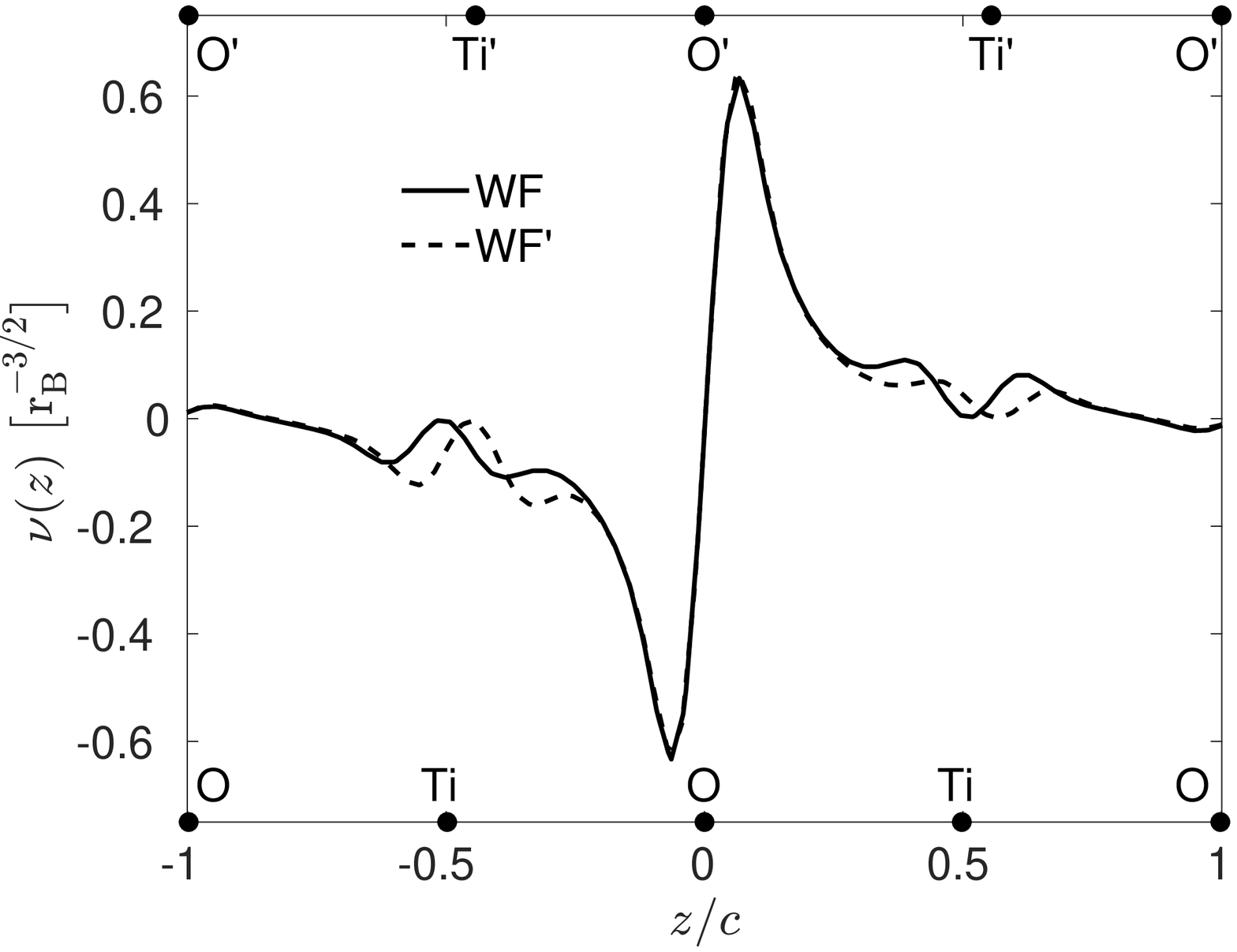}
\caption{Line plot along the $[001]$ direction of $\mathrm{O1}(p_z)$ WFs in the paraelectric and ferroelectric states of \BTO{}.
The $z$ coordinate is with respect to the $\mathrm{O}$ atom in the center.
The positions of the atoms in the paraelectric and ferroelectric states are depicted at the
bottom and top axes, respectively.
\label{fig_BTO_pz_cut}}
\end{figure}

Figure~\ref{fig_BTO_pz} displays the changes of the $\mathrm{O1}(p_z)$ WF going from a paraelectric phase ($P=0$) to the ferroelectric
equilibrium state ($P=P_s$). As evidenced by Fig.~\ref{fig_BTO_pz_para}, this orbital clearly shows a hybridization between
the Oxygen $p_z$ atomic orbital in the center of the figure and the $d_{z^2}$ atomic orbitals on the neighboring $\mathrm{Ti}$ atoms.
It thus forms a $\sigma$-type of bond oriented along the $\mathrm{Ti}$--$\mathrm{O}$--$\mathrm{Ti}$ chain.
This bonding changes in the ferroelectric state, as shown in Fig.~\ref{fig_BTO_pz_ferro}. Compared to the paraelectric state,
the hybridization strengthens for the lower $\mathrm{O}$--$\mathrm{Ti}$ bond and weakens for the upper one.
These modifications of the chemical bonding are due to electronic charge transfers induced by the atomic displacement.

\begin{figure}[!htbp]
\subfloat[paraelectric\label{fig_BTO_px_para}]{%
  \includegraphics[width=0.22\textwidth,clip=true]{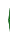}
}
\hfill
\subfloat[ferroelectric\label{fig_BTO_px_ferro}]{%
  \includegraphics[width=0.22\textwidth,clip=true]{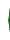}
}
\colorcaption{Amplitude isosurface plots of the $\mathrm{O1}(p_x)$ WFs in \BTO{} at $0.02~\mathrm{r_B}^{-3/2}$.
(a) Paraelectric state at $P=0$. (b) Ferroelectric state at $P=P_s$. For the location of the atoms
and the color scheme, see Fig.~\ref{fig_BTO_pz}.
\label{fig_BTO_wf_px}}
\end{figure}

\begin{figure}[!htbp]
\subfloat[paraelectric\label{fig_BTO_pxcut_para}]{%
  \includegraphics[width=0.22\textwidth,clip=true]{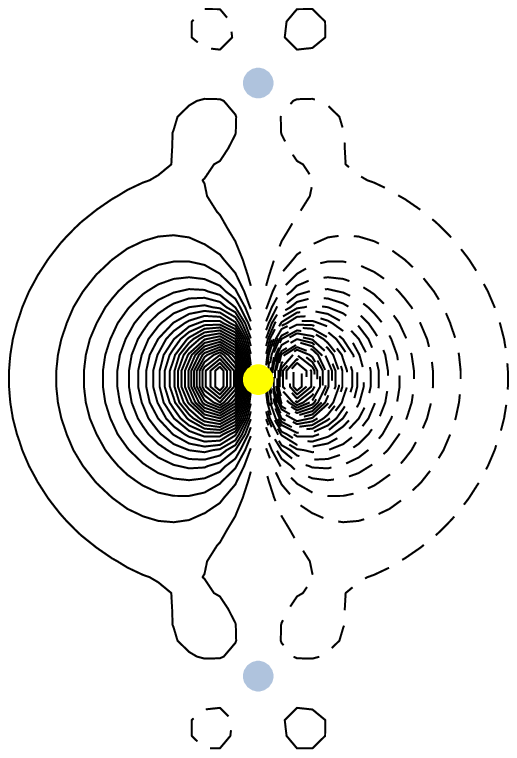}
}
\hfill
\subfloat[ferroelectric\label{fig_BTO_pxcut_ferro}]{%
  \includegraphics[width=0.22\textwidth,clip=true]{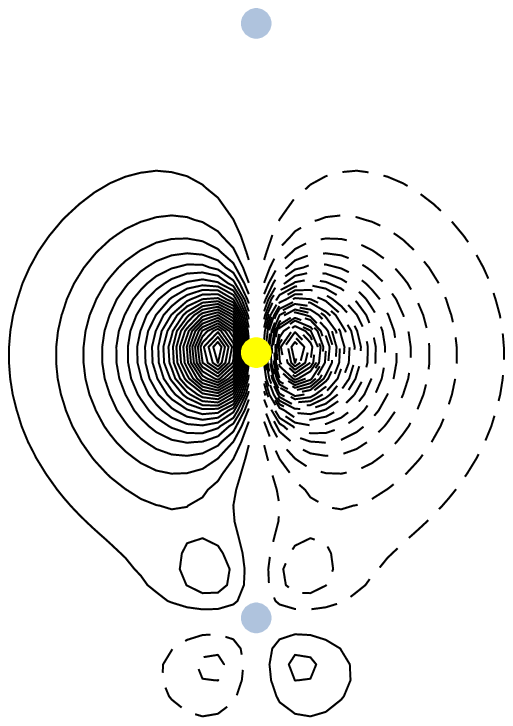}
}
\colorcaption{Contour plots in the $(020)$ plane of the $\mathrm{O1}(p_x)$ WFs in \BTO{}.
The contour intervals are $0.02~\mathrm{r_B}^{-3/2}$.
\label{fig_BTO_wf_pxcut}}
\end{figure}

To better visualize this process, we plot in Fig~\ref{fig_BTO_pz_cut} the overlayed cross-sections of the $\mathrm{O1}(p_z)$
WF along the $\mathrm{O}$--$\mathrm{Ti}$--$\mathrm{O}$--$\mathrm{Ti}$--$\mathrm{O}$ atomic chains in the paraelectric and
ferroelectric states. The amplitude of the orbital wave function increases in the ferroelectric state
around the left $\mathrm{Ti}$ atom, which is displaced from its initial position at $z$=$-0.5c$ towards the $\mathrm{O}$
atom in the center. This happens at the expense of the wave function around the right $\mathrm{Ti}$ atom, initially at $z$=$0.5c$,
which moves apart from the central $\mathrm{O}$ atom and whose amplitude decreases. Consequently, the hybridization to the
$\mathrm{Ti}$ $d_{z^2}$ states strengthens for shortened bond (Fig.~\ref{fig_BTO_pz} bottom), giving it a more covalent character,
and weakens for the elongated one (Fig.~\ref{fig_BTO_pz} top), resulting in a more ionic-like bond.
Note that the central part of the wave function does not change when the atoms move and retains its $p_z$ shape around the central
$\mathrm{O}$ atom. Thus, it can be concluded that the charge redistribution of the $\mathrm{O1}(p_z)$ orbital, as induced by the
atomic displacement, is non-local and due to off-site changes of hybridization at the neighboring $\mathrm{Ti}$ atoms.
The transfer of charge between the $\mathrm{Ti}$ atoms results in the displacement of the $\mathrm{O1}(p_z)$ orbital
centroid of charge, giving rise to electronic polarization.

A similar conclusion can be drawn by analyzing the transformations of the $\mathrm{O1}(p_x)$ and $\mathrm{O1}(p_y)$ WFs.
These orbitals form $\pi$-type bonds between the $\mathrm{O}$ and its neighboring $\mathrm{Ti}$ atoms,
as shown in Fig.~\ref{fig_BTO_wf_px} for the $\mathrm{O1}(p_x)$ WF. The form of the displayed wave function in the
paraelectric state, Fig~\ref{fig_BTO_px_para}, clearly shows the hybridization between the $p_x$ atomic orbital on the
$\mathrm{O}$ atom in the center and the $d_{xy}$ atomic orbitals on the neighboring $\mathrm{Ti}$.
This hybridization significantly changes in the ferroelectric state. As can be seen in Fig.~\ref{fig_BTO_px_ferro}
the admixing of $\mathrm{Ti}$ $d_{xy}$ contribution to the Wannier function gets much stronger for the bottom $\mathrm{Ti}$ atom,
while it almost disappears for the top one. At the same time the bulk part of the wave function in the form of a $p_x$ orbital
on the central $\mathrm{O}$ atom remains unchanged under the ferroelectric distortion. This can be better visualized
in Fig.~\ref{fig_BTO_wf_pxcut} which plots the contours of the $\mathrm{O1}(p_x)$ WFs projected onto the $(020)$ plane passing
through the \mbox{$\mathrm{Ti}$--$\mathrm{O}$--$\mathrm{Ti}$} chain along the $[001]$ axis parallel to the plane.
As is apparent from this figure, the transfer of charge takes place from the upper to the lower $\mathrm{Ti}$ atom, leaving the
central part of the wave function unaffected. Hence, similarly as for the $\mathrm{O1}(p_z)$ orbital, the mechanism responsible
for the anomalous displacement of the $\mathrm{O1}(p_x)$ orbital centroid of charge induced by the ferroelectric atomic distortion
is identified to be an interatomic transfer of charge between neighboring $\mathrm{Ti}$ atoms due to modified hybridizations.

The hybridization between the $p$ orbitals of $\mathrm{O}$ and $d$ orbitals of $\mathrm{Ti}$ in \BTO{} is a well-known feature,
confirmed by various sources: experiments, \cite{ref_BTO_exp_ElecSt_I,ref_BTO_exp_ElecSt_II}
calculations based on linear combination of atomic orbitals (LCAO), \cite{ref_BTO_lcao_I,ref_BTO_lcao_II,ref_BTO_lcao_III}
and DFT results. \cite{ref_Cohen_BTO_PTO,ref_btoWf}
Within the framework of DFT, Cohen and Krakauer \cite{ref_Cohen_BTO_PTO} deduced the increased hybridization between the
$\mathrm{O}$ $2p$ and $\mathrm{O}$ $3d$ states caused by the ferroelectric distortion by analyzing the densities-of-states
in tetragonal \BTO{} at experimental atomic displacements.
Marzari and Vanderbilt \cite{ref_btoWf} obtained maximally localized Wannier functions (MLWFs) in cubic \BTO{} from the postprocessing
step after a conventional electronic-structure calculation. Changes in hybridization were illustrated by manually displacing the
$\mathrm{Ti}$ atom along the \mbox{$\mathrm{Ti}$--$\mathrm{O}$} bond. It is worth mentioning that the qualitative
features of MLWFs are similar to our WFs. However, at variance with other approaches our method allows to directly inspect
the changes in hybridization and at the same time correlate it with the underlying energetics as the localized orbitals
are obtained by the requirement of the energy minimum.

\subsection{Enhanced ferroelectricity in tetragonal $\mathbf{PbTiO_3}$\label{sec_PTO}}

In this section, we use our approach to study the ferroelectricity of tetragonal \PTO{}.
The obtained results are compared to those of \BTO{}, discussed in the previous section,
to highlight the differences between both compounds.

\begin{figure}[!htbp]
\centering
\includegraphics[width=0.99\linewidth]{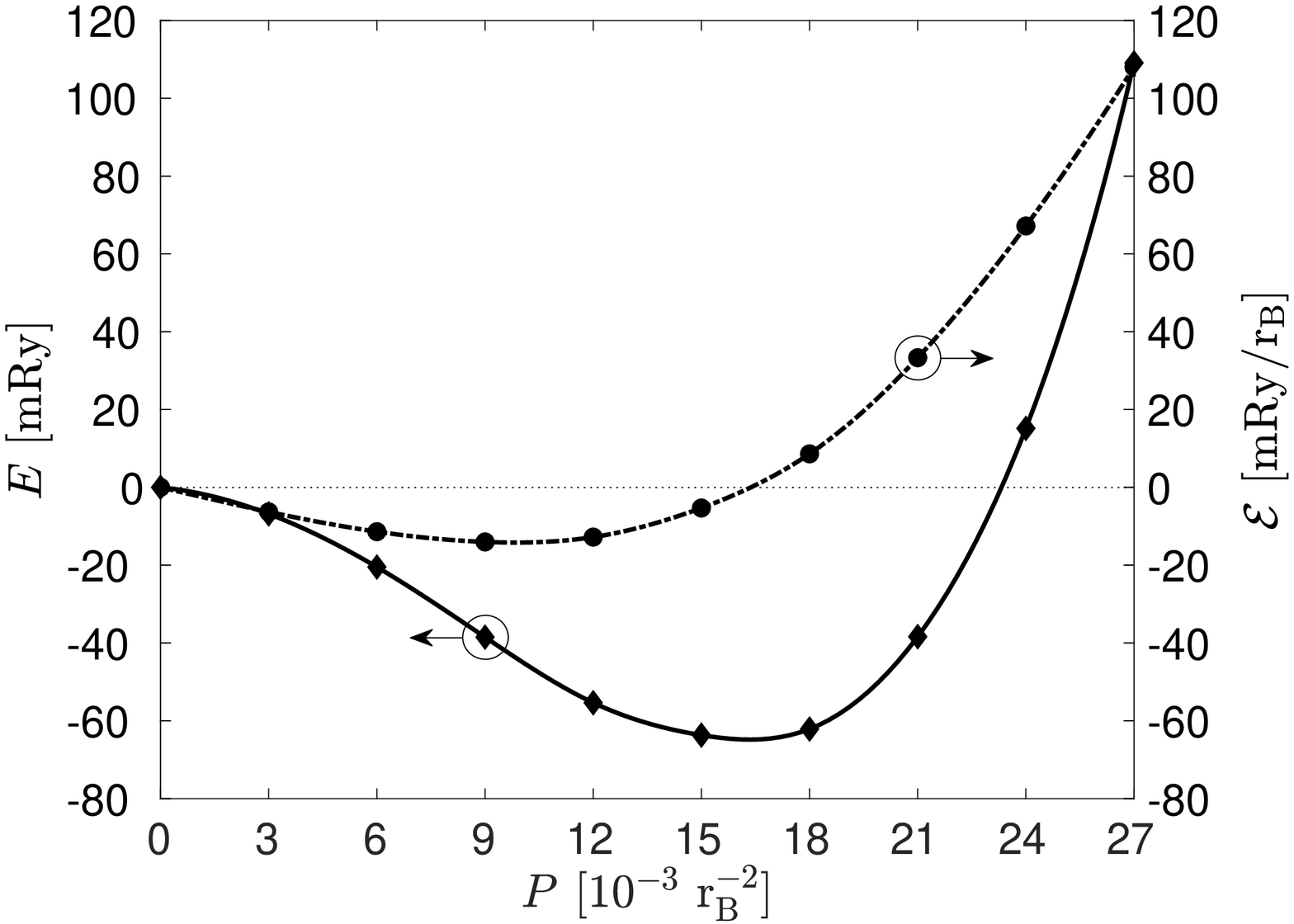}
\caption{Computed energy (diamonds) and electric-field (circles) as a function of the polarization for tetragonal \PTO{}.
The Kohn-Sham total energy $E$ is per $5$-atom unit cell. The electric field $\mathcal{E}$ and the polarization $P$
are the non-zero components along the $c$ axis.
\label{fig_PTO_Efield}}
\end{figure}

The calculated potential energy $E(P)$ curve and the values of the $\mathcal{E}$-field along the $c$ axis that
are required to impose the polarization constraint in tetragonal \PTO{} are plotted in Fig.~\ref{fig_PTO_Efield}.
Our calculations show that this compound stabilizes at a larger value of the spontaneous polarization,
$P_s=15\times10^{-3}~\mathrm{r_B}^{-2}$, and with a deeper potential energy well, $E_{\mathrm{min}}=63.7\mathrm{mRy}$,
as compared to \BTO{}. The larger derivatives of $E$ with respect to $P$ in the case of \PTO{} results in
higher values of the electric field, as expected from Eq.~(\ref{eq_enthalpy}). Similarly to \BTO{}, the zero-crossing of
$\mathcal{E}$ corresponds to the minimum location of $E$, thus verifying the internal consistency of the results.
The extracted value of $P_s$ for \PTO{} in SI units is $0.86~\mathrm{C}/\mathrm{m}^2$, which agrees reasonably well
with the experimental value \cite{ref_PTO_exp_pol} $0.75~\mathrm{C}/\mathrm{m}^2$. Since a similar conclusion
was drawn for \BTO{}, this gives a first hint about the well captured relative ferroelectric behavior of \BTO{} and \PTO{}
in our calculations. Another confirmation is provided by the large difference between the calculated well depths of
\BTO{} and \PTO{}, which was also reported in previous DFT studies \cite{ref_Cohen_BTO_PTO,ref_ferro_xc} using experimental
atomic displacements to investigate the potential energy surfaces of these materials.

\begin{figure}[!htbp]
\centering
\includegraphics[width=0.99\linewidth]{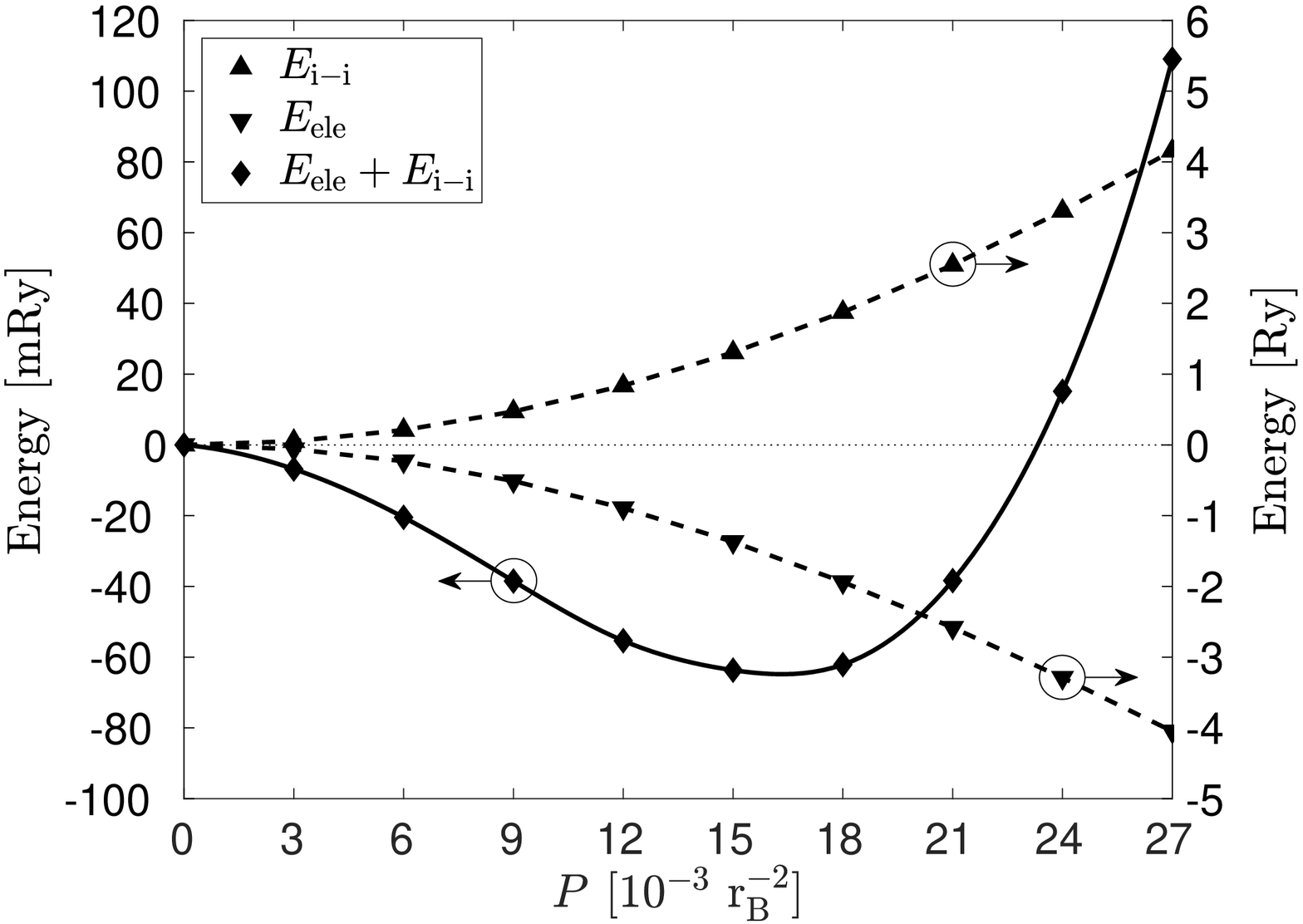}
\caption{Decomposition of the total energy in \PTO{} into the electronic, $E_{\mathrm{ele}}$, and ion-ion, $E_{\mathrm{i-i}}$, terms.
\label{fig_PTO_Edecomp}}
\end{figure}

Figure~\ref{fig_PTO_Edecomp} shows that the greater potential energy well depth for \PTO{} is due to enhanced
electronic processes, which give rise to a lowering of the electronic energy and help to stabilize the ferroelectric
state at larger spontaneous polarization relative to \BTO{}. This result also implies that the hybridization processes
should be stronger in the case of \PTO{}, a point that will be confirmed in the following.

\begin{figure}[!t]
\centering
\includegraphics[width=0.99\linewidth]{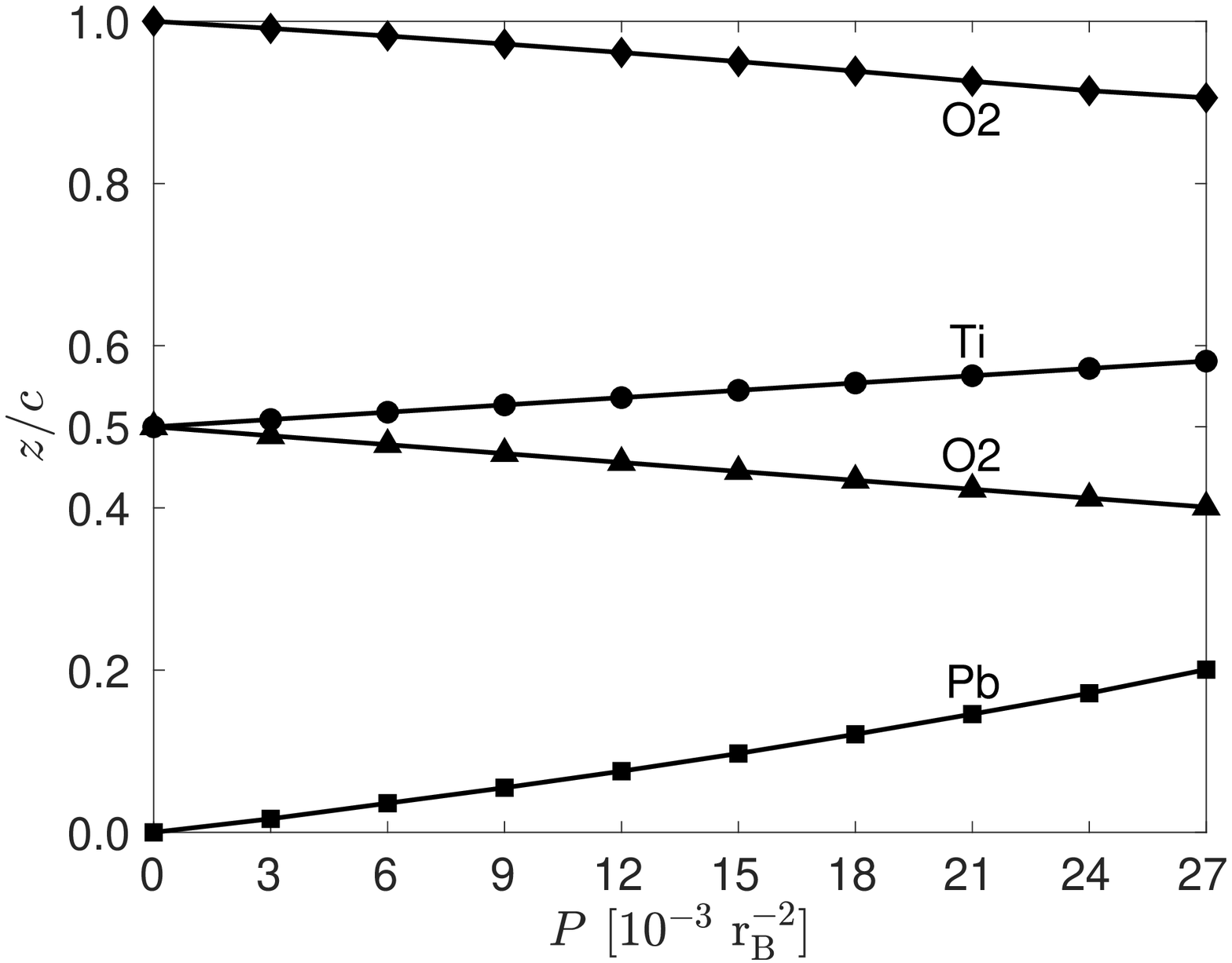}
\caption{Internal $z$ coordinate of each atom in the \PTO{} unit cell as a function of the polarization $P$
along the $c$ axis. The coordinates at $P=15\times10^{-3}~\mathrm{r_B}^{-2}$ correspond to the equilibrium
positions of the atoms.
\label{fig_PTO_zat}}
\end{figure}

The large polarization in \PTO{} is correlated with the substantial atomic distortions displayed in Fig.~\ref{fig_PTO_zat}.
As can be seen, the pattern resembles that of \BTO{}, for greater overall magnitudes of the
displacements of the corresponding atoms. A more detailed analysis of the structural distortions reveals however
that the ordering of the atoms according to the magnitude of their displacements with respect to their initial
positions is actually reversed in \PTO{} as compared to \BTO{}, i.e. $\{ \mathrm{Ti}, \mathrm{O1}, \mathrm{O2}, \mathrm{Pb} \}$
instead of $\{ \mathrm{Ba}, \mathrm{O2}, \mathrm{Ti}, \mathrm{O1} \}$. This result is supported by experimental data.
The measured fractional displacements of the $\mathrm{Ti}$, $\mathrm{O1}$, and $\mathrm{O2}$ atoms with respect to $\mathrm{Pb}$,
are \cite{ref_PTO_exp_struct} $(-0.049,-0.117,-0.120)$, whereas those extracted from our calculations at
$P_s$ are $(-0.052,-0.147,-0.152)$. Both simulation and experiment show that in \PTO{} the displacement of
$\mathrm{O2}$ towards $\mathrm{Pb}$ is greater than that of $\mathrm{O1}$ towards $\mathrm{Ti}$, while in \BTO{}
the situation is the opposite as the $\mathrm{O2}$--$\mathrm{Ba}$ displacement is smaller than the $\mathrm{O1}$--$\mathrm{Ti}$
movement (see~previous section). Thus, the difference in the character of the structural distortion between
\BTO{} and \PTO{} is well reproduced by our calculations.

In order to clarify the microscopic mechanism leading to large stabilized ferroelectric polarization in \PTO{}
relative to \BTO{} we will now turn to the inspection of the \PTO{} WFs and compare the electronic
structures of both materials.

The $\mathrm{O2}$ orbitals in \PTO{} respond to the atomic distortions alike the corresponding orbitals in \BTO{}.
The relative displacement of the OCs centered initially on the $\mathrm{O2}$ atomic sites is negligible when the
$\mathrm{O2}$ atoms move. An average of $-8$ electron charges can be associated with each $\mathrm{O2}$ site.
Summed up with the $+6$ charge of the $\mathrm{O}$ ion this gives a $-2$ effective charge contribution to the
total polarization coming from the displacement of the $\mathrm{O2}$ atoms, as in \BTO{}.

\begin{figure}[!b]
\centering
\includegraphics[width=0.99\linewidth]{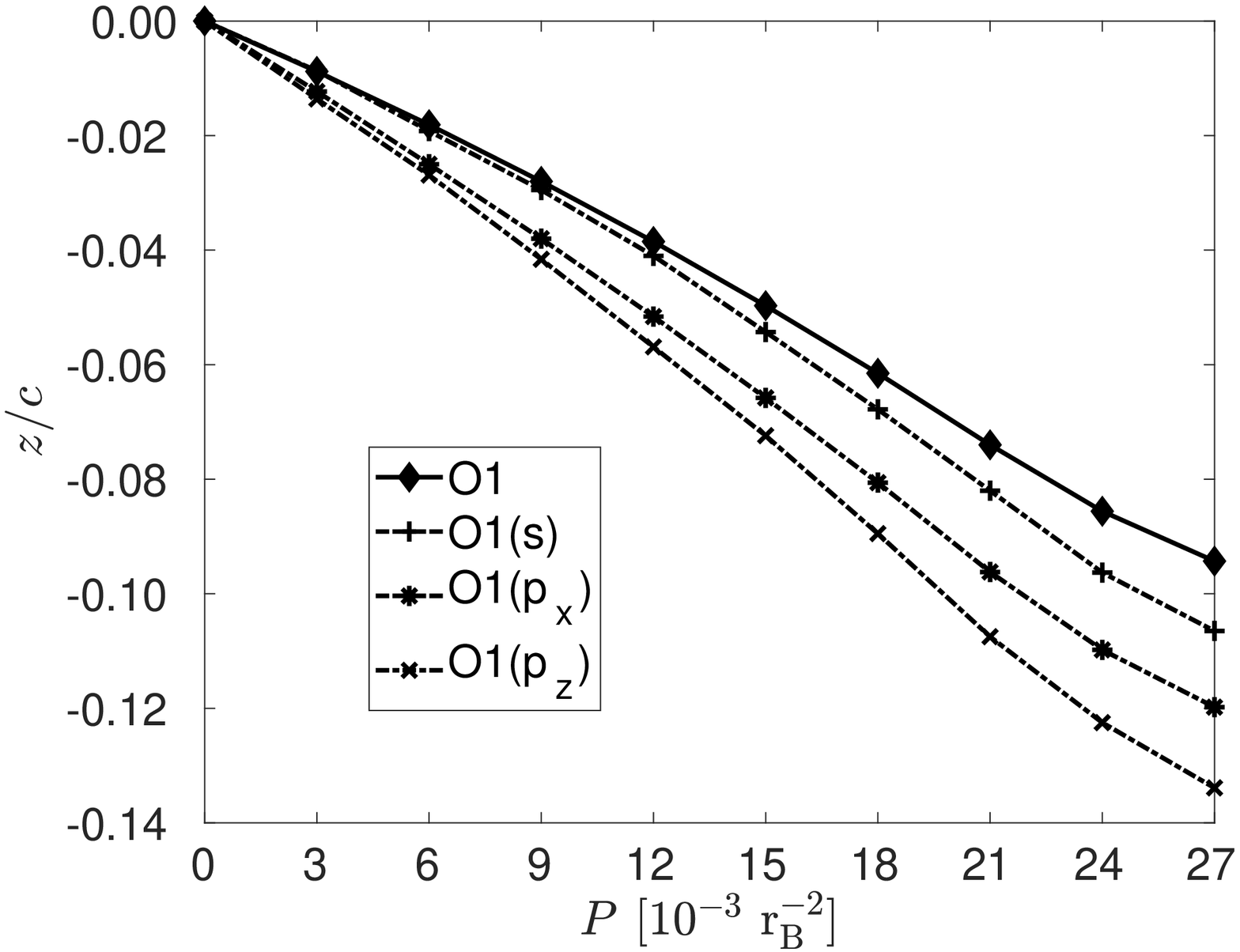}
\caption{Computed $z$ coordinate of the $\mathrm{O1}$ OCs (dashed lines) and of the $\mathrm{O1}$
atom (solid line) in \PTO{} as a function of the polarization $P$ along the $c$ axis.
The coordinates of the $\mathrm{O1}(p_y)$ OC overlap with those of the $\mathrm{O1}(p_x)$ OC.
\label{fig_PTO_zwf}}
\end{figure}

The larger stabilized polarization of \PTO{} can be partially attributed to the greater relative displacements of the
$\mathrm{O1}$ OCs with respect to the $\mathrm{O1}$ atom. The movement of the $\mathrm{O1}$ OCs and of the $\mathrm{O1}$
atom along the $c$ axis is displayed in Fig.~\ref{fig_PTO_zwf}. By comparing it to Fig.~\ref{fig_BTO_zwf}, which plots the
corresponding data for \BTO{}, it is evident  that the relative displacements of the $\mathrm{O1}$ OCs with respect to the
$\mathrm{O1}$ atom towards the $\mathrm{Ti}$ atom, are more significant in \PTO{} than in \BTO{}. As a consequence a larger
positive contribution to the electronic polarization comes from these orbitals in \PTO{}. A possible explanation is given
by the stronger hybridization between the $\mathrm{O}$ and $\mathrm{Ti}$ atomic orbitals in this compound. We find that the
character of the hybridization is the same as in \BTO{}. It is related to interatomic transfer of charge between the neighboring
$\mathrm{Ti}$ atoms. The difference in the displacement of the $\mathrm{O1}$ OCs and the related contribution to the electronic
polarization come from the stronger amplitudes of these processes in \PTO{} than in \BTO{}.

\begin{figure}[!htbp]
\centering
\includegraphics[width=0.99\linewidth]{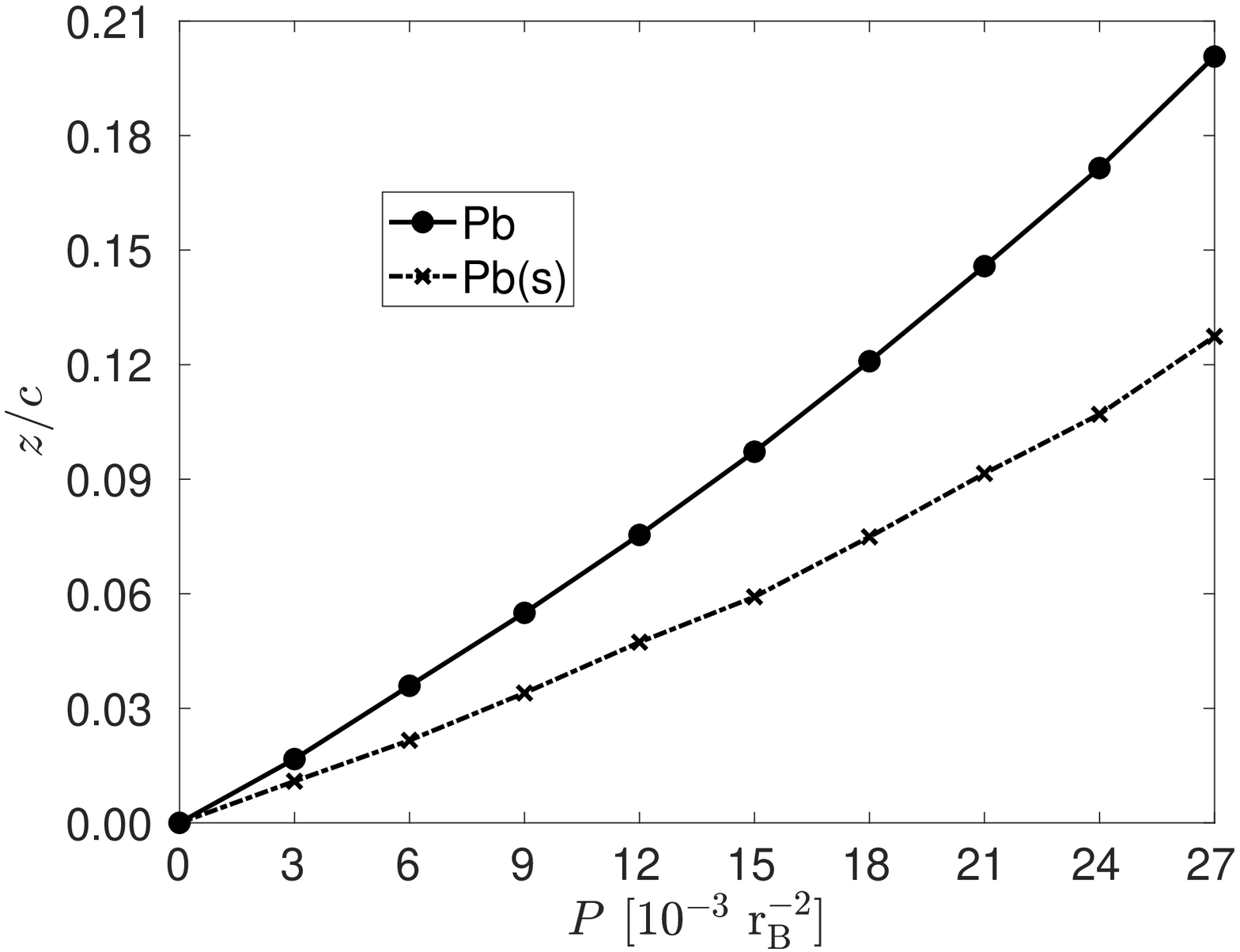}
\caption{Computed $z$ coordinate of the $\mathrm{Pb}(s)$ OC (dashed lines) and of the $\mathrm{Pb}$
atom (solid line) in \PTO{} as a function of the polarization $P$ along the $c$ axis.
\label{fig_PTO_zwf_Pb}}
\end{figure}

Another reason behind the relatively large polarization of \PTO{} is the presence of the $\mathrm{Pb}$ lone electron pair.
In our calculations these electrons are represented by a~doubly-occupied WF, which in the centrosymmetric structure is centered on the
$\mathrm{Pb}$ atom and displays a dominant atomic $s$ character. It is thus labeled as $\mathrm{Pb}(s)$. The~displacement of the
$\mathrm{Pb}(s)$ OC as a function of the polarization along the $c$ axis is shown in Fig.~\ref{fig_PTO_zwf_Pb}. As can be seen, the
$\mathrm{Pb}(s)$ OC lags behind the moving $\mathrm{Pb}$ atom. This leads to a larger positive contribution to the total polarization
than if the $\mathrm{Pb}(s)$ orbital would rigidly follow the $\mathrm{Pb}$ atom. In a purely ionic picture, where the $\mathrm{Pb}$
electrons remain centered on the $\mathrm{Pb}$ atom when moving, the contribution to the total polarization coming from the
displacement of the $\mathrm{Pb}$ ion would be +2 point charges, as the $\mathrm{Ba}$ ion in \BTO{}. This is not the case in \PTO{}.
Since the center of charge of the $\mathrm{Pb}(s)$ orbital is displaced downward with respect to the moving upward $\mathrm{Pb}$,
the positive charge of the $\mathrm{Pb}$ ion is exposed and a larger positive contribution to the total polarization than in a
purely ionic scenario is obtained.

\begin{figure}[!htbp]
\subfloat[paraelectric\label{fig_PTO_Pbs_para}]{%
  \includegraphics[width=0.22\textwidth,clip=true]{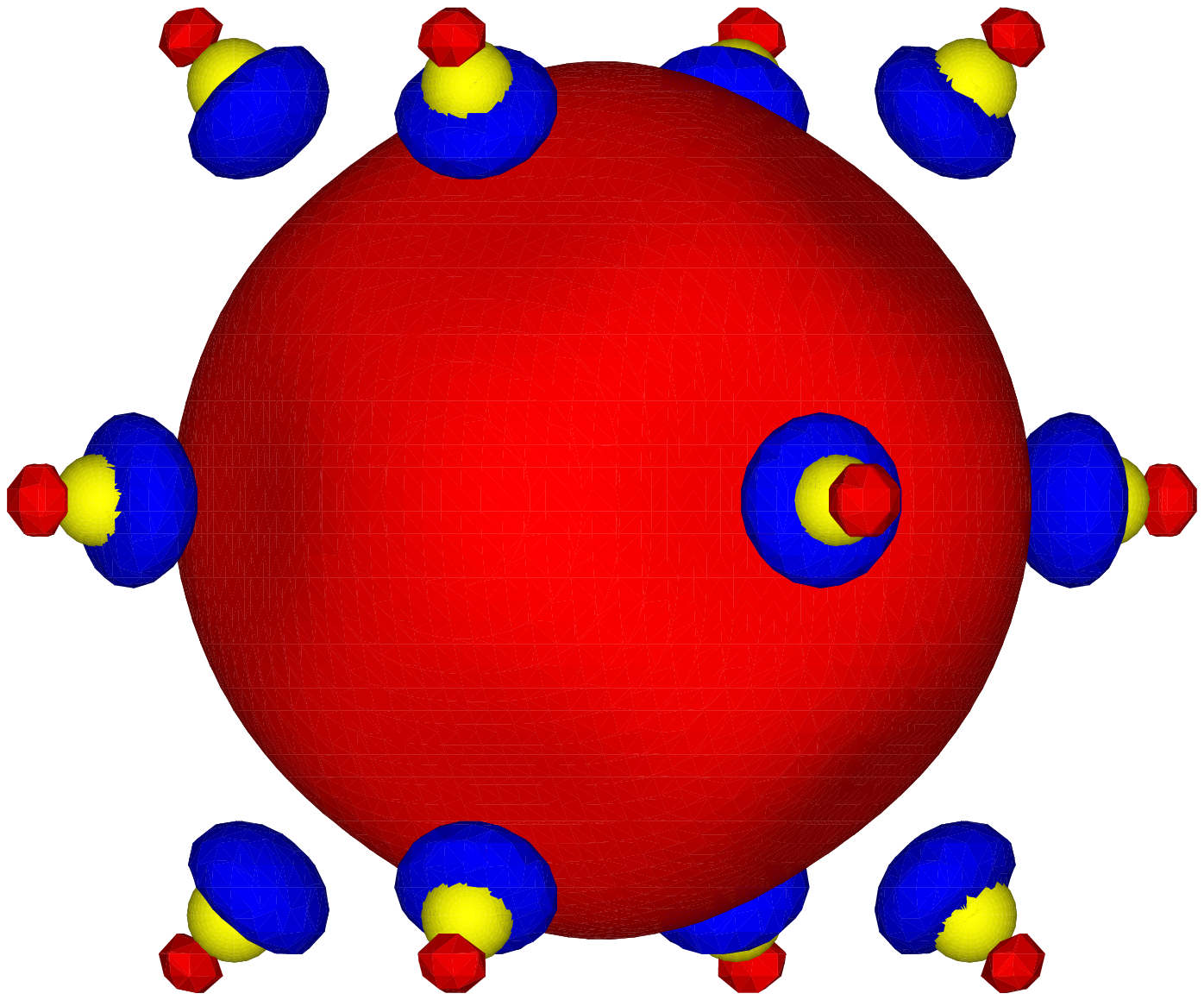}
}
\hfill
\subfloat[ferroelectric\label{fig_PTO_Pbs_ferro}]{%
  \includegraphics[width=0.22\textwidth,clip=true]{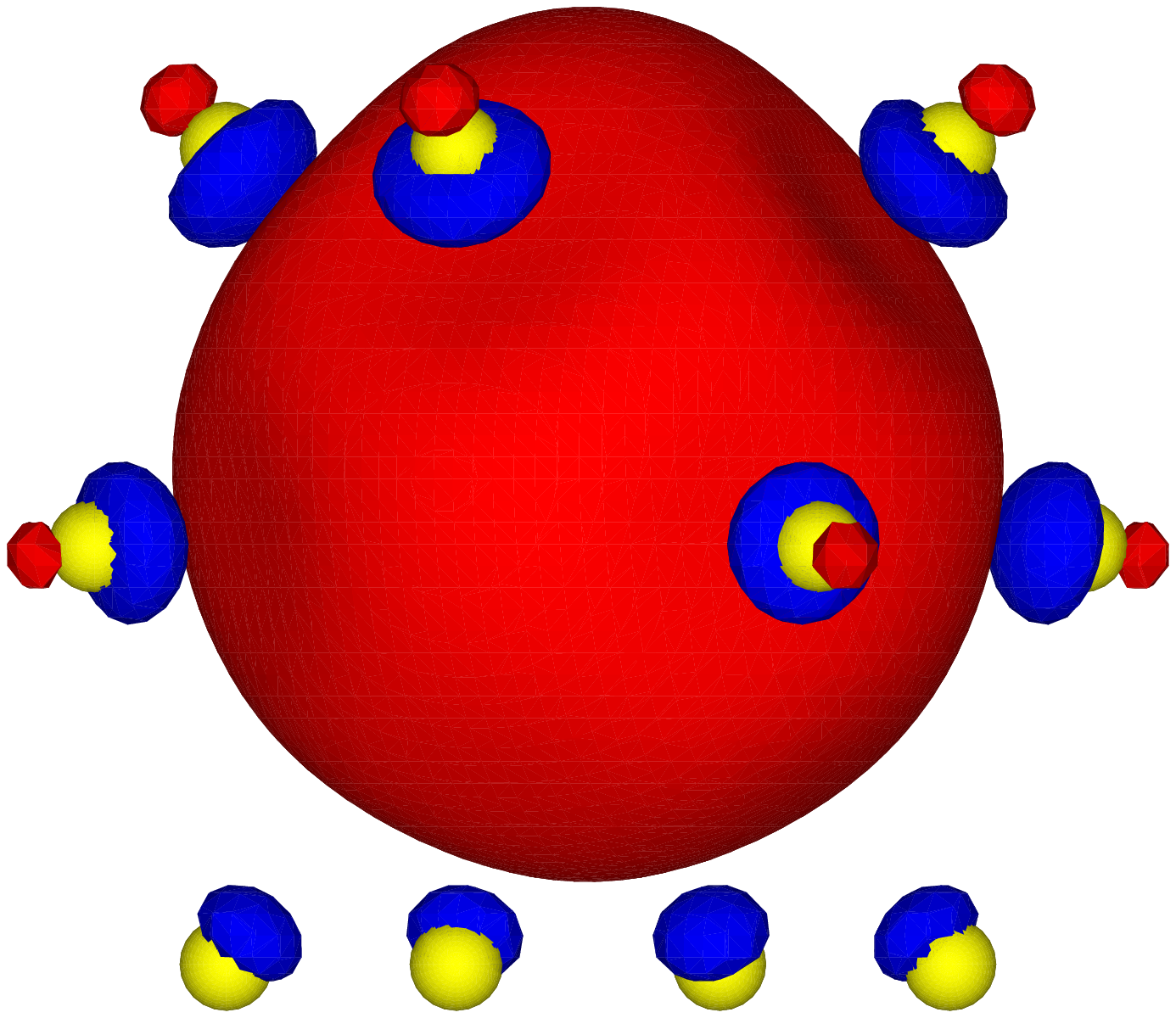}
}
\colorcaption{Amplitude isosurface plots of the $\mathrm{Pb}(s)$ WFs in \PTO{} at $\pm0.015~\mathrm{r_B}^{-3/2}$.
The red and blue surfaces correspond to positive and negative amplitudes, respectively.
$\mathrm{Pb}$ is at the center, hidden under an $s$ atomic orbital, surrounded by $12$ $\mathrm{O}$ atoms (yellow).
(a) Paraelectric state at $P=0$. (b) Ferroelectric state at $P=P_s$.
\label{fig_PTO_Pbs}}
\end{figure}

\begin{figure}[!htbp]
\centering
\includegraphics[width=0.99\linewidth]{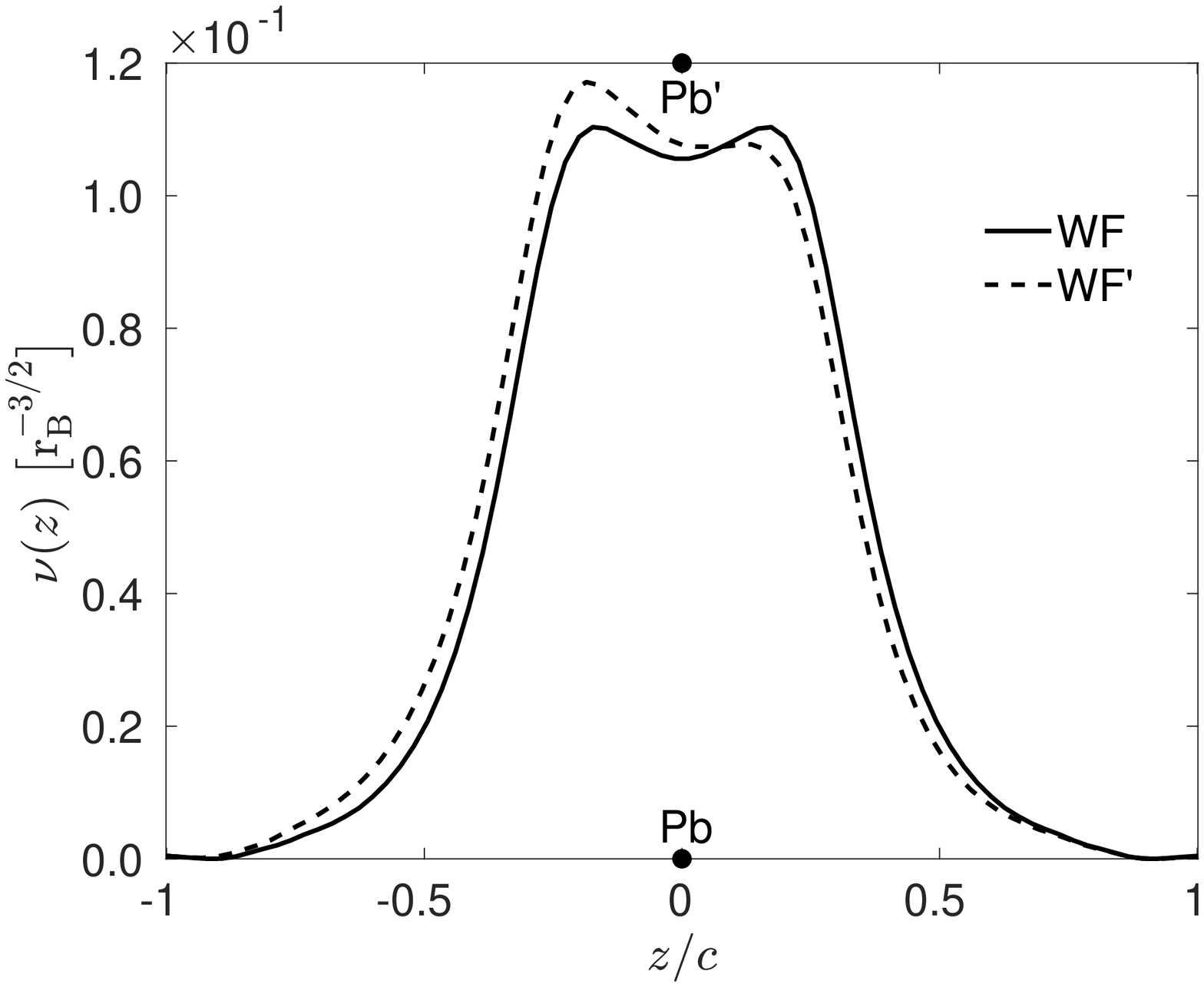}
\caption{Line plot along the $[001]$ direction of the $\mathrm{Pb}(s)$ WFs in the paraelectric and ferroelectric states of \PTO{}.
The $z$ coordinate is defined with respect to the $\mathrm{Pb}$ atom in the center.
\label{fig_PTO_Pbs_cut}}
\end{figure}

The mechanism responsible for the enlarged polarization contribution of the $\mathrm{Pb}$ atom, relative to its nominal
ionic charge, resembles a local electronic polarizability at the $\mathrm{Pb}$ site, as displayed in Fig.~\ref{fig_PTO_Pbs}.
In contrary to the $\mathrm{O1}$ orbitals, the $\mathrm{Pb}(s)$ shell charge entirely moves with respect to the $\mathrm{Pb}$
atom when transforming from the paraelectric to the ferroelectric state. This behavior is evident in Fig.~\ref{fig_PTO_Pbs_cut}.
The charge redistribution of the $\mathrm{Pb}(s)$ orbital in response to the atomic displacement is attributed to its interactions
with the neighboring $\mathrm{O}$ atoms. It can easily be seen in Fig.~\ref{fig_PTO_Pbs} that, in addition to the distinctive
atomic $s$ orbital on the central $\mathrm{Pb}$ site, there are significant $sp$-like contributions sitting on the 12 neighboring
oxygens. This supports the postulate that $\mathrm{Pb}$ in \PTO{} has a non-negligible covalent character.\cite{ref_Cohen_BTO_PTO}
In the ferroelectric state the $\mathrm{Pb}$--$\mathrm{O}$ hybridization increases for the upper $\mathrm{O}$ atoms and decreases
for the lower ones, resulting in interatomic charge transfers. However, the dominant mechanism driving the shift of the $\mathrm{Pb}(s)$
OC with respect to $\mathrm{Pb}$ atom is the on-site orbital reorganization following the change in the underlying crystal potential.
The later is caused by the relative displacements of $\mathrm{Pb}$ and $\mathrm{O}$ atoms.

The hybridization between the $\mathrm{Pb}$ $s$ and $\mathrm{O}$ $p$ states seems to indirectly influence the $\mathrm{Ti}$--$\mathrm{O}$ interactions,
which are mainly responsible for the ferroelectricity in both studied perovskite compounds. In addition, the favorable interactions between
$\mathrm{Pb}$ and $\mathrm{O}$ cause the energy to be lowered if the $\mathrm{Pb}$--$\mathrm{O2}$ distance is reduced. This leads to the conclusion
that it is the hybridization between $\mathrm{Pb}$ and $\mathrm{O}$ and the greater interactions between $\mathrm{Ti}$ and $\mathrm{O}$ that cause the
larger polarization and greater well depths of \PTO{}.

\section{Conclusion\label{sec_concl}}

We have presented a formalism to perform first-principles calculations of insulators at fixed polarization.
The approach has been implemented within the DFT framework into a practical computational
scheme that allows to find the most stable electronic and structural configuration of an insulating
crystal when its electric polarization is constrained to a given value. The method has been
applied to obtain the $E(\vec{P})$ curves for two paradigmatic ferroelectric materials, \BTO{} and
\PTO{} in their tetragonal phase.

In addition to qualitative results, the Wannier-function-based description of the electronic structure
of our approach yields a meaningful picture in real-space of the flow of electronic charge in terms of bonding.
It has been demonstrated how a careful analysis of the Wannier functions can shed light on relevant physics
concerning  the electronic polarization of perovskites. Two different basic mechanisms of electronic polarization,
interatomic charge transfers and local polarizability, have been visualized and associated with ferroelectric
transformations.

The comparative study of \BTO{} and \PTO{} using our approach has enabled us to clarify the special role of
$\mathrm{Pb}$ at the \mbox{$A$-site}. The presence of a $\mathrm{Pb}$ lone electron pair causes a
strong covalency between $\mathrm{Pb}$ and $\mathrm{O}$, resulting in larger polarization in \PTO{}
as compared to \BTO{}. Hybridization between $\mathrm{Pb}$-cation also leads to increased
$\mathrm{Ti}$--$\mathrm{O}$ interactions which further stabilize the ferroelectric state.
In both compounds the $\mathrm{Ti}$--$\mathrm{O}$ hybridization is crucial to allow for ferroelectricity.
With~the help of localized Wannier functions these processes can be directly inspected and quantified by
examining the relative displacements of the centroids of charge of the orbitals with respect to the
moving atoms at fixed polarization.

\bibliography{ferro_WF.bib}

\end{document}